\documentstyle[aps]{revtex}
\include{epsf}

\begin{document}


\draft

\title{Critical gravitational collapse of a perfect fluid:
nonspherical perturbations}

\author{Carsten Gundlach}

\address{Enrico Fermi Institute, University of Chicago, 5640 S. Ellis
Avenue, Chicago, IL 60637}

\address{Faculty of Mathematical Studies, University
of Southampton, Southampton SO17 1BJ, United Kingdom
\footnote{current address}}

\date{9 October 2001}

\maketitle


\begin{abstract}

Continuously self-similar (CSS) solutions for the gravitational
collapse of a spherically symmetric perfect fluid, with the equation
of state $p=\kappa\rho$, with $0<\kappa<1$ a constant, are constructed
numerically and their linear perturbations, both spherical and
nonspherical, are investigated. The $l=1$ axial perturbations admit an
analytical treatment. All others are studied numerically. For
intermediate equations of state, with $1/9<\kappa\lesssim 0.49$, the
CSS solution has one spherical growing mode, but no nonspherical
growing modes. That suggests that it is a critical solution even in
(slightly) nonspherical collapse. For this range of $\kappa$ we
predict the critical exponent for the black hole angular momentum to
be $5(1+3\kappa)/3(1+\kappa)$ times the critical exponent for the
black hole mass. For $\kappa=1/3$ this gives an angular momentum
critical exponent of $\mu \simeq 0.898$, correcting a previous
result. For stiff equations of state, $0.49\lesssim \kappa<1$, the CSS
solution has one spherical and several nonspherical growing modes. For
soft equations of state, $0<\kappa<1/9$, the CSS solution has 1+3
growing modes: a spherical one, and an $l=1$ axial mode (with $m=-1,0,1$).
\end{abstract}




\tableofcontents

\section{Introduction}

An isolated system in general relativity ends up in one of three
stable final states: a black hole, a star, or complete dispersion. The
phase space of isolated systems in general relativity is therefore
divided into basins of attraction: each initial data set must end up
in one of the stable end states. The study of the boundaries between
the basins of attraction, in particular of the boundary between the
black hole and dispersion end states, began with the pioneering work
of Choptuik \cite{Choptuik}, and is now an active field in classical
general relativity.

Initial data near the black hole/dispersion threshold evolve through a
universal intermediate state before dispersing or forming a black
hole. This intermediate attractor has higher symmetry, as a spacetime,
than the generic solution. Often it is self-similar. Close to the
threshold, but on the collapse side, the mass of the final black hole
then scales as a universal power of the distance of the initial data
to the black hole threshold. Universality, self-similarity and the
critical exponent for the black hole mass have given rise to the name
(type II) ``critical phenomena in gravitational collapse''. For a
review see \cite{critreview}.

Critical phenomena of this type are explained by the existence of a
solution that is self-similar, regular, and has exactly one growing
perturbation mode, such that for one sign of the growing mode the
solution veers towards black hole formation, and for the other towards
collapse. Such a solution is called a (type II) critical
solution. From a dynamical systems point of view, a critical solution
is an attractor within the black hole threshold, which is a
hypersurface of codimension one. Within the complete phase space, it
is therefore an attractor of codimension one. All solutions that start
near the black hole threshold, but not necessarily near the critical
solution itself, are funneled through this intermediate
attractor. This funneling process explains both universality and the
self-similar nature of the intermediate attractor explains
scaling. The critical exponent in the power-law scaling of the black
hole mass can be shown to be the inverse of the Lyapunov exponent of
the critical solution's one growing perturbation mode \cite{KHA1}.

Here we concentrate on one class of matter models coupled to general
relativity, perfect fluids with the linear barotropic equation of
state $p=\kappa\rho$, where $p$ is the pressure, $\rho$ is the total
energy density measured in the rest frame, and $\kappa$ is a constant
in the range $0<\kappa<1$. The spherically symmetric fluid with
$\kappa=1/3$, corresponding to an ultrarelativistic gas, was one of
the first systems in which critical phenomena were found
\cite{EvansColeman}. These results were later extended to the range
$0<\kappa<1$ \cite{Maison,KHA2,NeilsenChoptuik}. We now ask what
happens when we allow small deviations from spherical symmetry.

For a sample of values of $\kappa$ in the range $0<\kappa<1$, we
construct a regular, continuously self-similar (CSS), and spherically
symmetric solution, and then investigate its linear perturbations to
see how many growing perturbation modes it has. It is already known,
from both perturbative \cite{Maison,KHA2} and non-perturbative
\cite{NeilsenChoptuik} calculations that these solutions have exactly
one growing mode among their spherically symmetric perturbations,
which makes them critical solutions in spherical symmetry. Here we
examine their nonspherical perturbations. In a previous Rapid
Communication \cite{codim} we examined the particular case
$\kappa=1/3$. Here we generalize this investigation to all values of
$\kappa$ in the range $0<\kappa<1$. We also describe our numerical
methods and results in much more detail. Finally, we correct two
incorrect assumptions in \cite{codim}, namely that the $l=1$ axial
perturbations obey the same type of equation as the $l\ge2$ ones, and
that the $l=1$ polar perturbations are pure gauge. As it happens,
correcting these errors does not affect the overall conclusion of
Ref.~\cite{codim}, namely that the critical solution for $\kappa=1/3$
has no growing nonspherical perturbation modes.

As the background is spherically symmetric, the perturbations can be
separated into spherical harmonics labelled by $l$ and $m$, and can be
split further into axial and polar parts. The perturbation equations
we use are those derived in \cite{fluidpert}, restricted here to a CSS
background and the particular equation of state
$p=\kappa\rho$. Surprisingly, the nonspherical perturbation equations
are much harder to solve numerically than in the equivalent problem
for the massless scalar field \cite{critscalar}. One difficulty is
that we are dealing with a one-parameter family of background
solutions, whose limits $\kappa=0$ and $\kappa=1$ are not regular
members of the family. The other difficulty is that in the $l\ge 2$
perturbations both light cones and sound cones play a dynamical role,
while they coincide for the scalar field. This gives rise to weak
solutions (in the sense of hyperbolic equations) that we need to
discard. Because of these problems, our final choice of numerical
approach is to discretize the perturbation evolution equations in
space but not in time. We then look directly for eigenvectors and
eigenvalues (modes) of the finite difference equations.

The plan of the paper is this: in section \ref{section:background} we
discuss the general perturbation framework of Ref. \cite{fluidpert},
and the CSS background solutions. Section \ref{section:axial1}
discusses the $l=1$ axial perturbations. Their spectrum can be
calculated in closed form. Based on this, we correct the value of the
angular momentum critical exponent stated in Ref. \cite{angmom}.  All
other perturbations require a numerical treatment and are discussed in
Section \ref{section:otherperturbations}. Details of the numerical
difficulties and numerical methods, however, are given in the
appendixes. Section \ref{section:results} summarizes our results.

\section{Background solution}
\label{section:background}

\subsection{Perturbations of a spherically symmetric perfect fluid}

We shall examine the linear perturbations of a spherically symmetric
and continuously self-similar (CSS) perfect fluid spacetime. For this
purpose we use the restriction to a CSS background of a general
framework for the perturbations of time-dependent spherically
symmetric perfect fluid solutions that was presented in
\cite{fluidpert}. In this formalism, the spacetime manifold is written
as the product $M=M^2\times S^2$, where $S^2$ is the 2-sphere and
$M^2$ is a 2-dimensional manifold, the ``$rt$-plane'', with a boundary
$r=0$.  The coordinates in in $M^2$ are denoted by $x^A$, and the
coordinates in $S^2$ by $x^a$. Coordinates in all of $M$ are
collectively denoted by $x^\mu$. The general spherically symmetric
metric becomes, in this notation,
\begin{equation}
\label{spacetimemetric}
g_{\mu\nu} \equiv {\rm diag}\left(g_{AB}, r^2 \gamma_{ab}\right),
\end{equation}
where $r^2$ is a scalar function on $M^2$, and $\gamma_{ab}$ is the
unit metric on $S^2$. The spherically symmetric perfect fluid
stress-energy tensor can be written in the same notation as
\begin{equation}
T_{\mu\nu} \equiv {\rm diag}\left(\rho\, u_A u_B+p\,  n_A n_B, p\, r^2
\gamma_{ab}\right),
\end{equation}
where $\rho$ and $p$ are the (total) energy density and pressure in
the fluid rest frame, $u^\mu=(u^A,0)$ is the fluid 4-velocity, and
$n^A$ is the outward pointing unit vector normal to $u^A$ in
$M^2$. The field equations in this framework are covariant in $M^2$.
The two extra dimensions in $S^2$ appear through the scalar $r$.

The perturbations of a spherically symmetric background decompose
naturally into polar and axial parity, and into spherical harmonic
angular dependencies, for different $l$ and $m$. The equations of
motions are the same for all values of $-l\le m\le l$, for given $l\ge
0$. The cases $l=0$, $l=1$ (polar and axial), and $l\ge 2$ (polar and
axial) are all qualitatively different and need to be treated
separately. Tensors in $M$, including the perturbations, are written
as products of tensors in $M^2$ with tensors in $S^2$. All necessary
tensors in $S^2$ are built from the scalar spherical harmonics
$Y_{lm}$ on $S^2$, their covariant (with respect to $\gamma_{ab}$)
derivatives, and the covariantly constant antisymmetric tensor
$\epsilon_{ab}$. The final equations for the perturbations are again
covariant equations for tensors on $M^2$. Their angular dependence
comes in through terms such as $l/r$ and $l(l+1)/r^2$.

In the next step, linear combinations of the perturbations are found
that do not change to linear order under infinitesimal coordinate
transformations, in either $M^2$ or $S^2$. The perturbation equations
can be rewritten in terms of these gauge-invariant perturbations
alone.

In a further step, all perturbation tensors in $M^2$ are split into
frame components with respect to the orthonormal frame
$(u^A,n^A)$. This ``scalarization'' replaces covariant derivatives of
tensors with partial derivatives of scalars. These derivatives are
also decomposed into their frame $\dot f\equiv u^Af_{,A}$ and
$f'\equiv n^A f_{,A}$. The perturbation equations are now scalar
equations written without reference to a particular coordinate
system. In this sense they are covariant, as well as linearly
gauge-invariant.  Note that the frame derivatives are not partial
derivatives, and do not commute. The advantage of using these
derivatives is that $\dot f\pm c_s f'$ are derivatives along radial
matter characteristics, and $\dot f\pm f'$ derivatives along the
radial light rays. On the other hand, some constraint-type
perturbation equations for $l=0$ and $l=1$ are most naturally written
using the derivative $D \equiv r\partial f/\partial r$ along polar
slices.

In a final step, we rescale the perturbation variables by
$l$-dependent powers of $r$ so that they are either $O(1)$ at the
origin and even functions of $r$, or else $O(r)$ and odd. The
equations are brought into first-order form by treating first
derivatives such as $\dot f$ and $f'$ as independent variables where
necessary.

In the remainder of this section, we introduce coordinates that are
adapted to a CSS background, and review how the background solution is
defined, and constructed numerically.

\subsection{Continuously self-similar background solution}

Although our perturbation variables are linearly gauge-invariant, we
have to coordinate system on the background. 
A standard choice of coordinates $x^A$ is to use $r$ as a coordinate
(radial gauge), and make the second coordinate $t$ orthogonal to it
(polar slicing). Then $g_{AB}$ takes the form
\begin{equation}
\label{trmetric}
g_{tt}=-\alpha^2, \qquad g_{rr}=a^2, \qquad g_{rt}=0,
\end{equation}
with $\alpha$ and $a$ functions of $r$ and $t$. Radial light rays are
governed by the combination $g\equiv a/\alpha$. There is a remaining
gauge freedom $t\to t'(t)$, and we fix it by setting $\alpha=1$ at
$r=0$ for all $t$.

Based on radial-polar coordinates, we now introduce coordinates that are
adapted to self-similarity, while retaining polar slicing. We define
new coordinates $x$ and $\tau$ by
\begin{equation}
\label{xtau}
r\equiv sxe^{-\tau}, \qquad t\equiv -e^{-\tau},
\end{equation}
with $s>0$ a constant.  We have assumed that $t<0$, and have chosen
signs so that $\tau$ increases as $t$ increases. Note that
$\tau\to\infty$ as $t\to 0_-$. Partial derivatives transform as
\begin{equation}
f_{,t} = e^\tau(f_{,\tau}+xf_{,x}), \qquad f_{,r}=s^{-1}e^\tau f_{,x}.
\end{equation}
The metric $g_{AB}$ in these coordinates becomes
\begin{equation}
g_{\tau\tau}=e^{-2\tau}(-\alpha^2+s^2x^2a^2), \qquad
g_{xx}=e^{-2\tau}s^2a^2, \qquad g_{\tau x}=-e^{-2\tau}s^2 x a^2.
\end{equation}
The spacetime is continuously self-similar (with homothetic vector
$-\partial/\partial \tau$) if and only if $\alpha$ and $a$ depend only
on $x$ but not on $\tau$. $\tau$ has two different interpretations. On
the one hand, it is a time coordinate in the sense that its level
surfaces are spacelike. But $-\tau$ is also the logarithm of spacetime
scale, in the sense that proper distances are proportional to
intervals $\Delta x$ and $\Delta \tau$ times a factor of
$e^{-\tau}$. In a self-similar spacetime, larger $\tau$ therefore
means structure on a smaller scale. The point $r=0$, $t=0$, or
$\tau=\infty$, is by construction a curvature singularity, unless the
spacetime is flat. 

In the solutions we consider here the matter is a perfect fluid with
density $\rho$ and pressure $p=\kappa\rho$, with $\kappa$ a
constant. This equation of state is the only one compatible with exact
self-similarity. We impose CSS in the metric by making the
ansatz $a=a(x)$ and $\alpha=\alpha(x)$. Imposing self-similarity on
the spacetime, we find from the Einstein equations that
$4\pi\rho=e^{2\tau} \bar\rho(x)$ and $V\equiv u^A r_{,A} / n^B r_{,B}=
V(x)$. Here $n^A\equiv -{\epsilon^A}_B u^B$ is the outward-pointing
unit spacelike vector normal to $u^A$. The constant $s$ that was
introduced above is now chosen so that the surface $x=1$ is a matter
characteristic. It is then the past sound cone of the singularity.

The background solution that we want to use is completely defined by
the assumptions of i) continuous self-similarity, ii) spherical
symmetry, iii) analyticity at the center $x=0$, and iv) analyticity at
the past sound cone $x=1$. The background equations resulting from the
CSS ansatz are given in Appendix \ref{appendix:bgeqns}. The CSS ansatz
reduces the two Einstein equations and two matter equations that are
needed in spherical symmetry to one algebraic equation for $a$ three
ordinary differential equations in $x$ for $\rho$, $V$ and $g\equiv
a/\alpha$. 

There are two boundary conditions for a CSS solution at $x=0$. The
gauge condition $\alpha(0)=1$ becomes $g(0)=1$. From regularity of the
matter velocity, CSS, and matter conservation one can derive that
\begin{equation}
\label{matter_regularity}
\lim_{x\to0}{V\over sgx} = -{2\over 3(1+\kappa)}.
\end{equation}
Imposing this is the regularity condition iii) at the center.  Note
that this limit would also hold for a CSS fluid in a flat
spacetime. 

One boundary condition at $x=1$ is the gauge condition
$D(1)=0$, which makes $x=1$ is the sound cone. This condition
determines the value of the constant $s$. The regularity condition iv)
at the sound cone is
\begin{equation}
\sqrt{\kappa}S_1(1)=S_2(1), 
\end{equation}
which is the vanishing of the term in the equations that is divided by
$D$ (see Appendix \ref{appendix:bgeqns}).

Once we have solved the boundary value problem in $0\le x\le 1$, we
can analytically continue the solution through $x=1$ (which is a
regular singular point of the equations) and then continue the
solution by evolving the ODEs to larger $x$. (In numerical terms,
analytic continuation is implemented by polynomial extrapolation.)  We
go to the light cone and a little beyond. The light cone is at a value
of $x$ that depends on $\kappa$. The ODEs are regular there.

Our numerical method for imposing analyticity at $x=1$ and $x=0$ is to
just impose the algebraic boundary conditions there, and to use
centered differences everywhere else, without using an explicit
power-law expansion about the singular points. The first (numerical)
derivative of the fluid density and velocity profiles obtained with
this method has a small discontinuity at $x=1$ that first appears at
$\kappa\simeq 0.7$ and increases with $\kappa$. Results for $\kappa\gtrsim
0.7$ therefore have a source of numerical error over and above the one
arising in the numerical evolution of the perturbations.

\section{Axial $l=1$ perturbations -- analytical treatment}
\label{section:axial1}

\subsection{Equation of motion}

In this section we discuss the axial $l=1$ perturbations of the
continuously self-similar perfect fluid critical solution. This leads
us, from general arguments given in \cite{angmom}, to a prediction
for the scaling of black hole angular momentum in critical
collapse. Note that this section differs from the rest of the paper in
presenting purely analytical results.

The axial $l=1$ perturbations contain a single matter degree of
freedom, and no gravitational waves. The gauge-invariant fluid
velocity perturbation, $\beta$, obeys the autonomous equation of
motion
\begin{equation}
\label{betaeqn}
(\beta r^2 \rho u^A)_{|A} = 0.
\end{equation}
This is just a transport equation along the background fluid flow. All
axial metric perturbations are encoded in a gauge-invariant scalar
$\Pi$. For $l=1$, $\Pi$ is obtained from $\beta$ by quadrature, and
Eq. (\ref{betaeqn}) describes the dynamics completely.  For $l\ge2$,
$\Pi$ obeys a wave equation with a source term proportional to
$\beta$. (In \cite{angmom}, it was incorrectly assumed that this is
true also for $l=1$.)

As we shall see now, the complete mode spectrum of $\beta$ can be
obtained analytically for all $l$. For $l=1$, we then have the
complete dynamics. For $l\ge 2$, the spectrum of $\beta$ is also known
analytically, but that of the homogeneous $\Pi$ modes must be
calculated numerically. Here we obtain the modes of $\beta$ for
general $l\ge1$, and restrict to $l=1$ at the end.

The perturbed fluid velocity is regular at $r=0$ if $\beta$ is
$O(r^{l+1})$ there. We therefore define a rescaled variable $\bar\beta
= r^{-(l+1)}\beta$ that is even in $x$ and generically nonzero at
$x=0$. The perturbed stress-energy tensor remains self-similar if
$\beta$ depends on $\tau$ as $e^{-\tau}$ at constant $x$ (for any
$l$). We therefore define a rescaled variable $\overcirc\beta =
e^{-l\tau}\bar\beta=e^\tau (sx)^{-(l+1)}\beta $. This final
variable obeys the equation
\begin{equation}
\label{betacirc}
\overcirc\beta_{,\tau} + \left(x+{V\over sg}\right) \overcirc\beta_{,x}
+\left\{-1+{\kappa\over 1+\kappa}\left[2+\left(x+{V\over
sg}\right)(\ln \bar \rho)_{,x}\right] + (l+1)\left(1+{V\over
gsx}\right)\right\}\overcirc\beta = 0. 
\end{equation}
This is of the form
\begin{equation}
\label{transport}
{\overcirc\beta}_{,\tau} + xA(x) {\overcirc\beta}_{,x} + B(x)
{\overcirc\beta} = 0,
\end{equation}
where $A$ and $B$ are regular, even, strictly positive functions of
$x$, so that $A(x)=A_0 + A_2x^2 +O(x^4)$, and similarly $B(x)=B_0 +
B_2x^2 +O(x^4)$.  We look for solutions ${\overcirc\beta}$ that are
regular even functions of $x$.

\subsection{Analytic calculation of the mode spectrum}
\label{section:analyticspectrum}

Using the method of characteristics, the general solution of
(\ref{transport}) can be
written as
\begin{equation}
\label{characteristics}
{\overcirc\beta}(x,\tau)=e^{-B_0\tau} \exp\left(-\int_0^x
{B(x)-B_0\over xA(x)}\, dx\right) \ F\left[x \exp\left( -\int_0^x
{A(x)-A_0\over xA(x)}\, dx\ -A_0\tau\right)\right],
\end{equation}
where $F(z)$ is a free function that is determined by the initial
data. Note that the two definite integrals exist and are $O(x^2)$. For
regular, even initial data, with ${\overcirc\beta}(x,0)=F_0+O(x^2)$,
we have $F(z)=F_0+O(z^2)$.

We can now read off that the late-time behavior of the solution
${\overcirc\beta}(x,\tau)$ as $\tau\to\infty$ is 
\begin{equation}
\label{latetime}
{\overcirc\beta}(x,\tau)=e^{-B_0\tau}  \ \left[F_0 + O(x^2) +
O(e^{-2A_0\tau})\right ]
\end{equation}
as $\tau\to\infty$. For generic regular initial data, $F_0$ does not
vanish, and the solution decays as $e^{-B_0\tau}$ at late times.  One
might have expected that the growth exponent depends on details of the
background, but in fact it depends only on the background at the
center. The physical reason for this is that the transport equation
transmits information only from the center outwards. For example, we
can see from (\ref{characteristics}) that the solution
$\overcirc\beta$ for initial data that vanish in a neighborhood of
$x=0$ is strictly zero at any fixed $x$ at sufficiently large $\tau$.

Surprisingly again, we can evaluate $B_0$ in closed form, even though
the background solution away from the center is known only
numerically. From matter conservation and the assumption of continuous
self-similarity we have the regularity condition
(\ref{matter_regularity}). When we also take into account that
$\bar\rho$ is an even function of $x$ and that $v$ is an odd function,
we find from (\ref{betacirc}) that
\begin{equation}
\label{lambda}
\lambda=-B_0={2(1-3\kappa)-(1+3\kappa)l\over 3(1+\kappa)}.
\end{equation}
This analytic result is in perfect agreement with the late
time behavior of numerical evolutions of generic initial data for
$\overcirc\beta$.

We now show that the mode spectrum is discrete. To look for modes, we
make the ansatz
\begin{equation}
\overcirc\beta(x,\tau) = e^{\lambda\tau} f(x)
\end{equation}
and obtain 
\begin{equation}
\label{barf}
x{df \over dx} = -{\lambda + B\over A} f
\end{equation}
By expanding this equation in powers of $x$ around $x=0$ and comparing
coefficients, we see that if $f(0)\ne 0$, and $f$ is to be a regular
even function of $x$, we must have $\lambda=- B_0$. This is precisely
the mode that dominates the late-time behavior (\ref{latetime}), and
we may call it $f_0(x)$. One obtains it as a power series in a
neighborhood of $x=0$, and then by solution of the ODE (\ref{barf}).
We normalize $f_0$ by setting $f_0(0)=1$.  We now subtract a suitable
multiple of this mode from the solution $\overcirc\beta$ to obtain
something that is $O(x^2)$ at $x=0$:
\begin{equation}
{\overcirc\beta^{(2)}}(x,\tau) 
\equiv \overcirc\beta(x,\tau) - F_0 e^{-B_0\tau} f_0(x)
\end{equation}
where $\overcirc\beta(x,0)=F_0+O(x^2)$ as before. By construction
${\overcirc\beta^{(2)}}(x,0)=O(x^2)$. Expanding
(\ref{characteristics}) to the next order, we have
\begin{equation}
\label{latetime2}
{{\overcirc\beta^{(2)}}}(x,\tau)=e^{-(B_0+2A_0)\tau} \ \left[F_2 + O(x^2) +
O(e^{-2A_0\tau}) \right] x^2
\end{equation}
We then obtain a mode $f_2(x)$ by solving (\ref{barf}) with
$\lambda=-(B_0+2A_0)$. We normalize it as $f_2(x)=x^2+O(x^4)$.
Continuing in this way, we can strip off a sequence of modes decaying
with $\lambda=-(B_0+2nA_0)$ for $n=0,1,2,\dots$. These modes are
$O(x^{2n})$ at the origin, and so form a complete basis for smooth
functions $\overcirc\beta$. Therefore the entire spectrum is discrete.

\begin{figure}
\epsfxsize=8cm \centerline{\epsffile{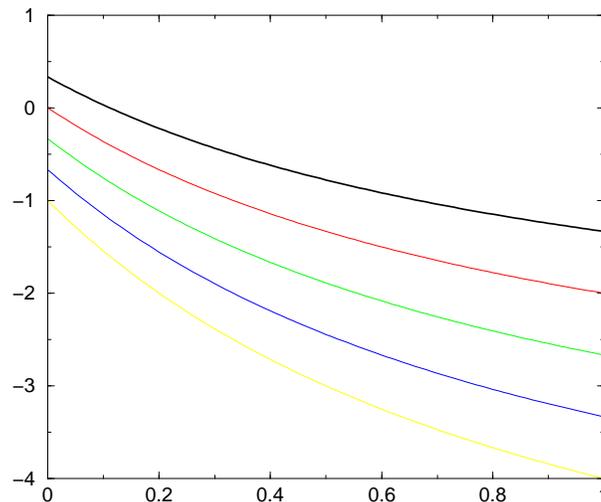}}
\caption{\label{figure:beta1-5} The $\overcirc\beta$ equation
describes axial fluid velocity perturbations. The plot shows the
growth exponent $\lambda$, which is real, against $\kappa$. We plot
$\lambda(\kappa,l,n)$ given in Eq. (\ref{betaspectrum}). The thick
line is for $l=1$ and $n=0$, where $\overcirc\beta$ is the only
perturbation. (It is positive for $\kappa<1/9$.) Below, from top to
bottom $l=2\dots5$, for the leading mode, with $n=0$.}
\end{figure}

From (\ref{betacirc}), using again (\ref{matter_regularity}), we find
\begin{equation}
A_0={1+3\kappa\over 3(1+\kappa)}.
\end{equation}
We can now write down the complete spectrum for all values of $\kappa$ and
$l$, labelled by the index $n=0,1,2,\dots$. It is
\begin{equation}
\label{betaspectrum}
\lambda(\kappa,l,n)=-B_0(\kappa,l)-2nA_0(\kappa)=
{2(1-3\kappa)-(1+3\kappa)(l+2n)\over
3(1+\kappa)}.
\end{equation}
From this formula we can read off that all $l\ge 2$ modes decay for
all $\kappa$ in the range $0<\kappa<1$. All $l=1$ modes also decay for
$\kappa >1/9$, but for $\kappa<1/9$ there is exactly one growing $l=1$
mode (the $n=0$ mode). $\lambda$ for the dominant ($n=0$) $l=1$ mode
is relevant for angular momentum scaling. It is
\begin{equation}
\label{lambda1}
\lambda_1={1-9\kappa\over 3(1+\kappa)}.
\end{equation}
The dominant ($n=0$) mode is plotted for
$l=1\dots5$ in Fig.~\ref{figure:beta1-5}.

\subsection{Angular momentum scaling}

In \cite{angmom} we derived a general formula for the critical
exponent $\mu$ governing black hole angular momentum in critical
collapse, namely
\begin{equation}
\label{mu}
\mu = {2-\lambda_1\over\lambda_0} = (2-\lambda_1)\gamma.
\end{equation}
(This formula corrects a misprint in Equation (11) of Reference
\cite{angmom}.) Here $\lambda_0$ is the Lyapunov exponent for the
spherical mode, which is real and positive, and $\lambda_1\equiv
\lambda(\kappa,1,0)$ is the Lyapunov exponent for the $l=1$ axial
perturbations, which is real. $\gamma=1/\lambda_0$ is the critical
exponent for the black hole mass.  The derivation of this formula
assumes that the critical solution has precisely one growing
perturbation mode, which is spherically symmetric, while all
nonspherical perturbation modes decay. The analytical result in this
section and the numerical results in the following sections show that
this assumption holds in the range $1/9<\kappa \lesssim 0.49$.

For this range of equations of state, putting the results
(\ref{lambda1}) and (\ref{mu}) together, we obtain the analytical
prediction
\begin{equation}
\label{muofk}
\mu(\kappa) = {5(1+3\kappa)\over 3(1+\kappa)} \gamma(\kappa), \qquad {1\over9}<\kappa\lesssim 0.49, 
\end{equation}
for the angular momentum exponent $\mu$, given the mass exponent
$\gamma$.

This result corrects the value of $\mu$ given in \cite{angmom} for the
case $\kappa=1/3$, based on an incorrect value of $\lambda_1$. The
correct value $\lambda_1=-1/2$ for $\kappa=1/3$ gives a real critical
exponent $\mu=5\gamma/2\simeq 0.898$ for the black hole angular
momentum.  Note that the incorrect value given in \cite{angmom} was
complex, which was expected to give rise to oscillations in the
direction of the black hole angular momentum as the black hole
threshold is approached. However, the correct value of $\lambda_1$ is
real, and so the angular momentum scaling is a power law, like the
mass scaling.

\section{All other perturbations - numerical treatment}
\label{section:otherperturbations}

\subsection{General aspects of the perturbation equations}

Before we discuss the perturbation equations of motion in detail, it
is useful to discuss the general form of the perturbations of a
spherically symmetric and CSS solution. As we have seen, we can choose
dependent and independent variables for the background so that the
background is given by a number of functions $Z(x,\tau)=Z_*(x)$ of a
single self-similarity variable $x$. It follows that the equations of
motion of the linear perturbations of this background, when written in
first-order form in suitable variables, are of the general form
\begin{eqnarray}
\label{evolution_equations}
u_{,\tau} &=& A(x)u_{,x} + B(x)u , \\
\label{constraint_equations}
u_{,x} &=& C(x)u.
\end{eqnarray}
Here $u(x,\tau)$ is a vector of perturbations, and $A$, $B$, $C$ are
matrices that depend on the background. We shall call the coupled
partial differential equations that contain $\tau$-derivatives
evolution equations, and the coupled ordinary differential equations
in $x$ constraints. Not all variables $u$ need obey both types of
equation -- some variables $u$ obey only an evolution equation, others
only a constraint equation, and yet others both. 
For most of the perturbations, we shall be able to use a free
evolution scheme in which the only constraints are trivial ones of the
form $u_{1,x}=u_2$ introduced by writing a wave equation in first-order form.
For the spherical perturbations and the polar $l=1$ perturbations,
however, it is unavoidable to solve nontrivial constraint equations. 

The perturbation equations
(\ref{evolution_equations},\ref{constraint_equations}) admit solutions
of the form
\begin{equation}
\label{modesolution}
u(x,\tau) = e^{\lambda\tau}u_\lambda(x) .
\end{equation}
Both $\lambda$ and $u_\lambda(x)$ are in general complex, but because
the coefficients $A$, $B$ and $C$ are real, the solutions form complex
conjugate pairs. With $\lambda\equiv\Lambda\pm i\omega$, 
the general real solution is
\begin{equation}
{\rm Re}[Ce^{i\delta}u(x,\tau)] =  Ce^{\Lambda\tau}\left[
\cos(\omega\tau + \delta)\,{\rm Re \,} u_\lambda(x) 
-\sin(\omega\tau + \delta)\,{\rm Im \,} u_\lambda(x)\right].
\end{equation}
If the general solution can be written as a sum over such modes is a
subtle question, but here we ask mainly if
there are any growing modes with $\Lambda>0$. Furthermore, if we
discretize $u(x,\tau)$ on a finite grid in $x$ (but not in $\tau$),
the general solution of the discretized field equations is then
clearly given by the sum over a finite number of modes, each of which
is exactly exponential in $\tau$.

In defining the variables $u$ that we evolve numerically, we begin
from the all the variables with an overbar defined in
\cite{fluidpert}, and their dot and dash derivatives where
necessary. On a self-similar background, it is useful to further
rescale these variables by powers of $e^\tau$ so that the background
plus perturbations is still self-similar if and only if the rescaled
variables are independent of $\tau$. If this is done, the resulting
equations do not contain explicit powers of $e^\tau$. The spacetime
perturbations grow in a physical sense towards the singularity
if and only if the variables $u$ grow with $\tau$. Rescaled barred
variables will be denoted by a circle, their rescaled dot derivatives
by a tilde, and their rescaled prime derivatives by a hat.

In the following, we shall use ``degree of freedom'' to denote a
variable that can be freely specified as a function of a radius at the
initial moment of time. In this count, a wave equation has two degrees
of freedom, for each value $l$ and $m$. In the $l\ge 2$ perturbations
there are eight physical degrees of freedom, corresponding to wave
equations for the two polarizations of gravitational waves, the three
components of the Euler equation, and the continuity equation. On a
spherically symmetric background, these generic eight degrees of
freedom split into three axial and five polar degrees of freedom. The
number of first-order variables is larger (four and seven,
respectively) because in the first-order form of a wave equation for a
variable $\phi$, $\phi$ itself and $\phi_{,x}$ are separate variables
that are linked by a (trivial) constraint equation.

\subsection{Numerical methods}

In investigating the non-spherical perturbations numerically, we have
to treat each value of $\kappa$ and $l$ separately. (The equations do
not depend on $m$.) In practice, we work with a finite sample of
values of $\kappa$ in the range $0<\kappa<1$, and with $l\le 5$. With
increasing $l$, numerical difficulties at the center of spherical
symmetry become more pronounced, limiting the range of $l$ we can
investigate. Fortunately the range $l\le 5$ is sufficient to see a
trend, as we shall demonstrate in plots.

We are looking for mode solutions of the form (\ref{modesolution}),
and in particular for the dominant mode, the one with the lowest value
of $\Lambda\equiv{\rm Re}\lambda$. This objective allows a number of
very different numerical approaches. Three different ones come to
mind, and we describe them first, and then summarize our experience
with the last two of them. More details are given in the appendixes.

1) Making the ansatz (\ref{modesolution}), and imposing suitable
regularity conditions, one obtains a boundary value problem for an
eigenvector $u_\lambda(x)$ and eigenvalue $\lambda$. For the $l\ge 2$
axial perturbations, regularity conditions have to be imposed at the
center and at the light cone. For the $l=1$ polar perturbations
regularity conditions have to be imposed at the center and the
soundcone, and for the $l\ge 2$ polar perturbations regularity
conditions are required at the center and both the light cone and the
soundcone (which is in the interior of the numerical domain). Modes
can then be found in two ways:

a) From an initial guess for $\lambda$ and $u_\lambda(x)$, one can
find the correct values by shooting or relaxation. In practice, the
initial guess has to be quite good, and finding one solution does not
exclude the possibility that there is another solution with larger
$\Lambda$.

b) For given $\lambda$, the boundary (and possibly midpoint)
conditions can be solved in terms of $n$ free parameters. If the
boundary value problem is well-posed, the shooting procedure must
match up $n$ variables. The mismatch in these $n$ variables is a
linear function of the $n$ free parameters, and is therefore described
by an $n\times n$ matrix $A$ that depends analytically on
$\lambda$. If $A(\lambda)$ has a kernel, a solution $u_\lambda$ can be
found for this $\lambda$. One therefore looks for zeros of the
complex-analytic function $\det A(\lambda)$. This can be done by
contour integrals \cite{Gundlach_Chop2}.

2) One can also use the equations to evolve generic initial data
$u(x)$ in $\tau$. At late times the solution will be dominated by the
dominant mode, and one can read off $\lambda$ from its
time-dependence. One can also subtract the dominant modes one after
another, in the Gram-Schmidt process , in order to find subdominant
modes. This is known as the Lyapunov method \cite{KHA2}.

3) In a third approach, the evolution equations are finite differenced
in $x$ but not in $\tau$. The resulting finite difference-differential
equations can be used in two ways:

a) With $M$ degrees of freedom on $N$ grid points in $x$, the map
$T:u_i(x_j)\to u_{i,\tau}(x_j)$ with $i=1\dots M$ and $j=1\dots N$ is
a square matrix of size $(MN)^2$. Its eigenvalues with the largest
real parts should be an approximation to the continuum eigenvalues
$\lambda$ with largest real part. (The lower eigenvalues will depend
on the discretization scheme and are not expected correspond to
continuum modes.)

b) Alternatively, we can use a standard ODE integration scheme to
discretize in time. Such a numerical method is called
``semi-discrete'' because with sufficiently small step size
$\Delta\tau$ it is effectively discrete only in $x$. The map
$T_\Delta:u(0,x)\to u(\Delta,x)$ for a finite interval $\Delta$ is
again an $(MN)^2$ matrix. The few eigenvalues with largest modulus
should now be approximations to the largest of the numbers
$e^{\Delta\tau}$.  

We first implemented the Lyapunov method, method 2). It is the
simplest method for obtaining the dominant mode. This method worked
well for the $\kappa=1/3$ fluid \cite{codim}, and also for the
nonspherical perturbations of the scalar field critical solution
\cite{critscalar}. However, in the polar perturbations for some values
of $\kappa$ and $l$, the dominant mode is a numerical artifact (an
instability) in all finite differencing schemes that we have
tried. (The nature of the instabilities will be discussed below.) Then
one needs to look for sub-dominant modes of the finite difference
equations in order to find the dominant physical mode. We have found
that the Lyapunov method is extremely inefficient for finding
subdominant modes. 

We then implemented both method 3a) and 3b). In method 3b) we used
first order, second order and fourth order Runge-Kutta integration
(RK1, RK2, RK4), and an implicit second-order scheme (iterated
Crank-Nicholson, or ICN). For a small enough time steps, the
differences between these methods are negligible, and we effectively
reach the continuum limit in $\tau$. Furthermore, in this limit, the
modes and eigenvalues produced by methods 3a) and 3b) agree up to
roundoff error. Therefore, there is no advantage in method 3b) over
3a), but a higher computational cost.

Within method 3), many ways of finite differencing in $x$ are
possible. We have used four finite differencing schemes in $x$. The
differences between them remain important at all feasible values of
$\Delta x$.  Two of these schemes are upwind schemes that explicitly
use the eigenvalues of the matrix $A$. Both are the linearized version
of Godunov schemes. GD1, the scheme used before in \cite{codim}, is
first-order accurate. GD2 is a second-order accurate version.  The
other two schemes used centered differences and are second-order
accurate. CD3 uses the obvious centered differences. CD4 uses a
well-known trick to deal with terms of the form
$\phi_{,x}+(2l/x)\phi$, which can give rise to numerical instabilities
near the center $x=0$. (The 3 and 4 are just consecutive labels.) All
four schemes are defined in Appendix \ref{appendix:methods}. 

Some of the numerical instabilities that we see are familiar: problems
at the center, in particular for high $l$, and grid modes in centered
differencing. Another kind of instability was harder to understand.
The continuum equations admit mode solutions (\ref{modesolution}) in
which $u_{,x}$ is discontinuous at a characteristic that is also a
line of constant $x$, that is, at the light cone and/or the sound cone
of the singularity. The evolution equations are hyperbolic, and these
solutions are called weak solutions. They are discussed in more detail
in Appendix \ref{appendix:weakmodes}. While they are valid as a
generalized type of solution, these modes would not arise in a
collapse situation, and so we need to exclude them. Unfortunately, for
certain values of $\kappa$ and $l$, they dominate the top smooth
physical mode.

The weak modes were not seen in the investigation \cite{critscalar} of
the perturbations of the scalar field critical solution, because there
they could be discontinuous only at the light cone of the singularity,
but the numerical domain was truncated precisely there. The same can
be done for the axial perturbations of the perfect fluid critical
solution, because there are no axial sound waves. Similarly, the
numerical domain can be truncated at the sound cone for the spherical
and $l=1$ polar perturbations because they do not comprise
gravitational waves. The $l\ge2$ polar perturbations, however, contain
coupled sound and gravitational waves. Therefore the numerical domain
cannot be truncated at smaller $x$ than at the light cone, and this
leaves weak modes at the sound cone. 

The finite differencing schemes we use are not designed to represent
weak solutions correctly, but they do of course have a counterpart in
the modes of the finite difference system. Sometimes the numerical
counterpart resembles the continuum mode (in particular in GD1 and
GD2), but sometimes it cannot be distinguished by inspection from a
smooth mode (in particular in CD3 and CD4). The only certain criterium
is convergence. This makes it crucial that we have more than one
finite differencing scheme, so that we can carry out independent
residual evaluations, as well as simple convergence tests.

We have loosely referred to the numerical artifacts as
instabilities. However, the usual concept of the stability of a
numerical method is at most exponential growth of numerical
solutions. But here we are using semi-discrete methods (discrete only
in $x$) on a system of linear equations with $\tau$-independent
coefficients. Therefore all solutions, both physical and artificial,
depend exactly exponentially on $\tau$. Exponential growth, or its
absence, can therefore not be used to distinguish between physical and
unphysical solutions. An unphysical mode may grow either more or less
rapidly than a physical mode. The only certain way of distinguishing
them is by convergence. (This will also rule out the weak modes,
because the numerical methods were not designed to handle them, and
will therefore fail to converge on them.)

We have carried out two kinds of convergence test. One test is to
check that the discretized eigenvectors $u_{\lambda}(x_j)$ and
corresponding eigenvalues $\lambda$ converge with increasing
resolution in $x$ at the expected rate (to first order or second
order) for each scheme, but also that all four schemes converge to the
same solution.  The other test is independent residual
evaluation. Writing the system of continuum equations formally as
$u_{,\tau}(x)=Lu(x)$, where $L$ is a linear derivative operator, let
$L_1$ and $L_2$ be two different finite difference approximations to
$L$, and let $(u_1,\lambda_1)$ and $(u_2,\lambda_2)$ be modes
(eigenvectors and eigenvalues) of $L_1$ and $L_2$. If these are
approximations to a continuum mode, the norms $|\lambda_1 u_1-L_2
u_1|$ and $|\lambda_2 u_2 - L_1 u_2|$ should converge to zero with
increasing resolution. Convergence should be to second order in
resolution if both methods are second-order accurate, and to first
order in resolution if one or both methods are only first-order
accurate. For the norm $||$ we choose the $l^2$-norm divided by the
number $N$ of grid points, which is an approximation to the $L_2$
norm.

Imposing the constraints (\ref{constraint_equations}) poses no
numerical difficulty. During evolution, in methods 2) or 3b), one has
the choice of either imposing the constraints only on the initial
data, from time to time, or at each time step. We find that this
hardly affects the results. Some care has to be taken when calculating
the map $T$ in the presence of constraints. This is discussed in
Appendix \ref{appendix:constraints}.

\subsection{Spherical perturbations}

The spherical perturbations have already been investigated by several
authors \cite{Maison,KHA2,NeilsenChoptuik}, and we use them here as a
test of our methods, and also to make sure that we are investigating
the same background solution as these authors. Gauge-invariant
perturbations do not exist for $l=0$. Equations of motion for the
spherical perturbations are most easily obtained by linearizing the
field equations in spherical symmetry in the polar-radial gauge that
is also used for the background. The perturbations $\delta(\ln\rho)$
and $\delta V$ obey evolution equations in $\tau$ and $x$, while the
perturbations $\delta a$ and $\delta g$ obey constraint equations in
$x$ only. We do not give the detailed equations here.

We have used the all four codes. In GD1, GD2 and CD3 the top physical
(growing) mode shows up as the top numerical mode. In CD4 the growing
physical mode is also present but must be identified by hand because
it is dominated by grid modes. 

\begin{figure}
\epsfxsize=16cm
\centerline{\epsffile{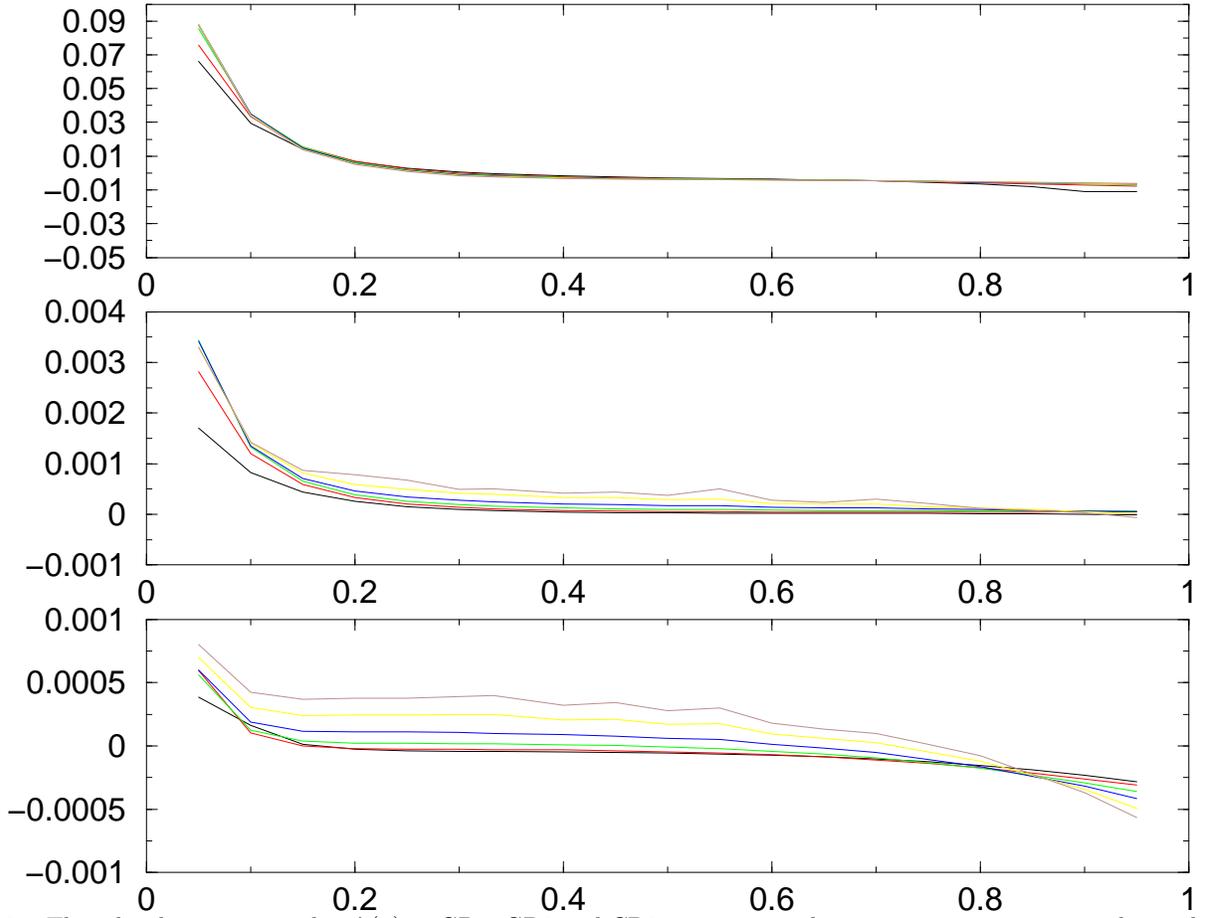}}
\caption{\label{figure:sph_conv} This plot demonstrates that
$\lambda(\kappa)$ in GD1, GD2 and CD3 converges with increasing to a
common value and at the expected order. The resolutions are $\Delta
x=1/10,1/20,...1/320$. The three plots, from top to bottom, show the
error in GD1, GD2 and CD3. A reference value for $\lambda(k)$ was
obtained by Richardson interpolation on CD3 at the highest three
resolutions. This reference value was then subtracted from the result
of all three codes at all resolutions to obtain a measure of error. To
demonstrate power-law convergence, this error was divided by a factor
of 4 for each factor of 2 in $\Delta x$ (for CD3 and GD2), or by a
factor of 2 (for GD1). The error at the highest resolution was not
rescaled. Note the rise of the error at low $\kappa$ in all three
methods. Although the different graphs in each plot do not lie on top
of each other perfectly, they are similar, indicating approximate
power-law scaling at the expected order.  }
\end{figure}

Fig. \ref{figure:sph_conv} demonstrates the convergence of GD1, GD2
and CD3 with increasing resolution towards a common value of
$\lambda(k)$ over the entire range of $\kappa$. GD1 converges
approximately to first order, and GD2 and CD3 approximately to second
order, as expected. The finite differencing error is approximately
$10^{-3}$ in both second-order schemes at $\Delta x=1/320$ for
intermediate values of $\kappa$. It rises towards the high and the low
end of the $\kappa$ range.

\begin{figure}
\epsfxsize=8cm
\centerline{\epsffile{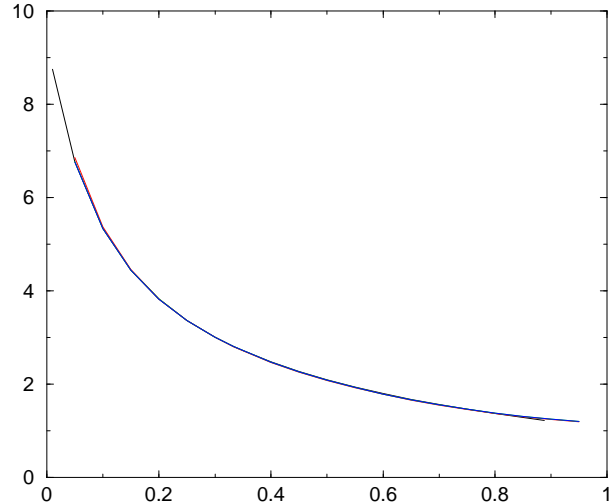}}
\caption{\label{figure:sph_maison_and_ic1} $\lambda(\kappa)$ from
Maison, and our codes GD1, GD2 and CD3 at $\Delta x=1/320$. (The four
lines are not resolved in this plot.)}
\end{figure}

\begin{figure}
\epsfxsize=8cm
\centerline{\epsffile{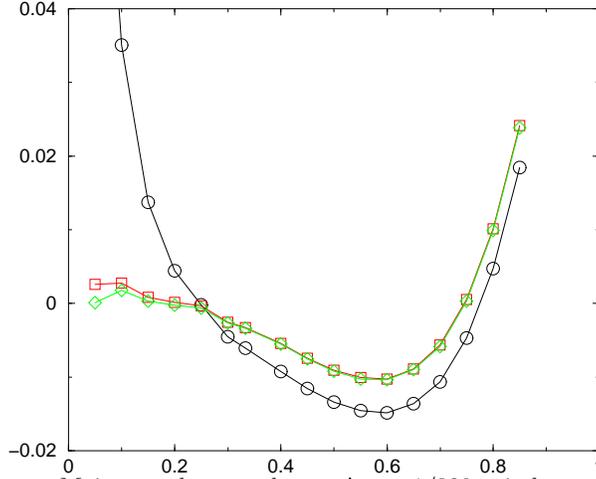}}
\caption{\label{figure:sph_maison_comp} Difference in $\lambda(k)$
between Maison and our codes at $\Delta
x=1/320$: circles are GD1, squares GD2 and diamonds CD3. The fact that
GD2 and CD3 differ from Maison by approximately the same value
indicates that this difference is mainly systematic error, rather than
finite differencing error. The fact that GD has a different deviation
from Maison is explained by its larger finite differencing error,
which increases at low $k$ -- see top plot of Fig. \ref{figure:sph_conv}.}
\end{figure}

Figs. \ref{figure:sph_maison_and_ic1} and
Fig. \ref{figure:sph_maison_comp} compare our $\lambda(k)$ at $\Delta
x=1/320$ (GD1, GD2 and CD3) with that obtained by Maison
\cite{Maison}. The results agree quite well for all $\kappa$.
Nevertheless, the figures show that there is a systematic difference
to Maison's results that is generally much larger than our finite
difference error. It grows in all three codes as $\kappa\to 1$. (The
exception is in GD1 at low $\kappa$ where the finite difference error
becomes dominant.) We suspect that the systematic error is in our
code, rather than Maison's code, and specifically in the background
code: as we have discussed above, it is not well-behaved at the light
cone as $\kappa\to 1$, and at the center as $\kappa\to 0$.

\subsection{Axial $l\ge2$ perturbations}
\label{section:axialge2}

For $l\ge2$ the gauge-invariant velocity perturbation $\beta$ obeys an
autonomous transport equation, just as for $l=1$. For $l\ge 2$, there
is also a gauge-invariant metric perturbation $\Pi$, which obeys a
wave equation with $\beta$ as a source \cite{fluidpert}. The evolution
of $\beta$ is still autonomous. This means that the $\lambda$-spectrum
of the coupled $\Pi$ and $\beta$ system is the sum of two parts: modes
in which $\beta$ vanishes, so that they are solutions of the $\Pi$
equation without a source term, and modes that are driven by $\beta$,
so that their value of $\lambda$ is set by the evolution of $\beta$. 

As shown above in section \ref{section:axial1}, the spectrum of of
$\beta$-modes can be calculated analytically for all $l$ including
$l\ge2$. As a check we have implemented the $\beta$ equation on its
own, and the numerical results agree with the analytical ones, showing
the top two of the analytically calculated modes. That is as much as
one can expect, because the $n$-th mode behaves at the center as
$x^{2n}$, and no finite differencing scheme is designed to represent
such behavior correctly. In fact, unphysical modes that behave as
$x^n$ with $n$ odd also show up. They are unphysical because the
corresponding velocity perturbation is not regular at the center, but
they are valid solutions of the equation.

The equations for $\Pi$ were derived in the fluid frame in
\cite{fluidpert}, but they look simpler in the frame of constant $r$
observers (radial frame). These are just different choices of
first-order variables for the same wave equation. As a test, we have
implemented the equations in both frames, and the results converge as
expected. In the fluid frame we have also implemented the coupling to
$\beta$. As expected this simply adds extra modes driven by $\beta$
modes to the spectrum. As we have analytical results for the
$\beta$-modes (they all decay), we only need the free $\Pi$ modes. 

For simplicity, we give here only the source-free equations in the
fluid frame. These equations are quite similar to the toy model wave
equation of Appendix \ref{section:toymodel}. The variables are
\begin{equation}
\overcirc\Pi \equiv e^{-l\tau} \bar\Pi, \qquad
\tilde\Pi \equiv e^{-(l+1)\tau} \alpha^{-1}\Pi_{,t}, \qquad
\hat\Pi \equiv e^{-(l+1)\tau} a^{-1}\Pi_{,r}.
\end{equation}
$\overcirc\Pi$ and $\tilde\Pi$ can be specified
freely on the initial surface, while $\hat\Pi$ is constrained by
\begin{equation}
\hat\Pi=as\overcirc\Pi_{,x}. 
\end{equation}
The $\Pi$ equation without source, in the radial frame, is then
equivalent to 
\begin{eqnarray}
\tilde\Pi_{,\tau} & = & A_1 \tilde\Pi_{,x} + B_1 \hat\Pi_{,x} 
-\left(l+1+{C_a\over sg}\right)\tilde\Pi
+{1\over sg}\left[{2(l+1)\over x}+C_\alpha\right]\hat\Pi
-(l+2){1\over s^2ag}\left[{C_g\over x}+(l-1){a^2-1\over
x^2}\right]\overcirc\Pi, \\
\hat\Pi_{,\tau} & = & C_1 \tilde\Pi_{,x} + A_1 \hat\Pi_{,x}
+{C_\alpha \over sg}\tilde\Pi - \left(l+1+{C_a\over sg}\right)\hat\Pi,
\\
\overcirc\Pi_{,\tau} &=&{a\over g}\tilde\Pi - sxa \hat\Pi - l\overcirc\Pi.
\end{eqnarray}
This is of the form $u_{,\tau}=Au_{,x}+Bu$, 
where the coefficients of the $2\times 2$ matrix $A$ are 
\begin{equation}
\label{lightcoeffs}
A_1 = -x, \quad B_1 = {1\over sg}, \quad C_1 = {1\over sg}.
\end{equation}
The matrix $A$ has therefore the eigenvalues
\begin{equation}
\lambda_{1\pm}=-x\pm {1\over sg}.
\end{equation}

\begin{figure}
\epsfxsize=8cm \centerline{\epsffile{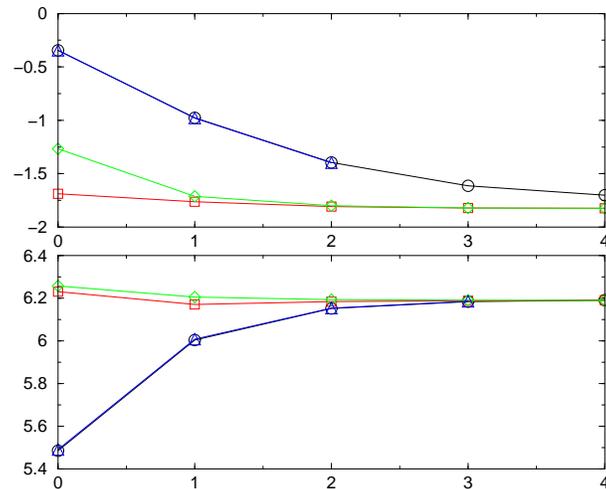}}
\caption{\label{figure:pi2_conv} Convergence of $\lambda$ for
the top physical $l=2$ axial ($\Pi$ only) mode with resolution, at
$\kappa=1/3$. The upper graph is ${\rm Re}\lambda$, and the lower graph is
${\rm Im}\lambda$. From left to right $N=10,20,\dots 160$. Circles are
GD1, squares are GD1LC, diamonds GD2 and triangles up CD3.}
\end{figure}

\begin{figure}
\epsfxsize=8cm \centerline{\epsffile{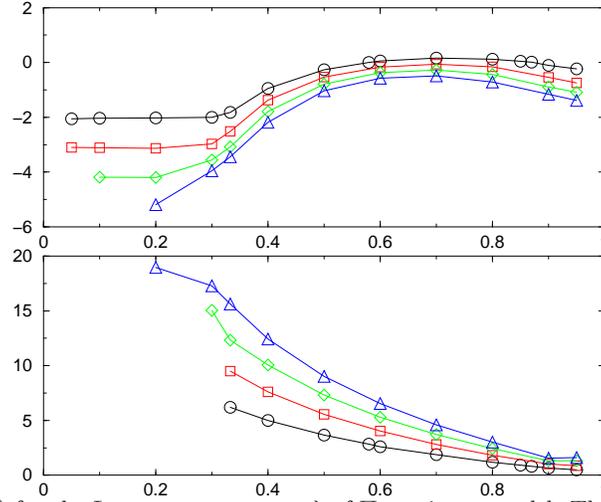}}
\caption{\label{figure:pi2-5} Best value (using CD4) for the Lyapunov exponent
$\lambda$ of $\Pi$, against $\kappa$ and $l$. The upper graph shows ${\rm
Re}\lambda$ against $\kappa$, and the lower graph ${\rm
Im}\lambda$. Circles denote $l=2$, squares $l=3$, diamonds $l=4$ and
triangles $l=5$. Points are linked by straight lines. The points
$\kappa=0.05,0.1$, $l=5$ and $\kappa=0.05$, l=4 are missing because $\lambda$
could not be computed. The curves for ${\rm Im}\lambda$ end at small
$\kappa$ where the modes become real.}
\end{figure}

\begin{figure}
\epsfxsize=8cm \centerline{\epsffile{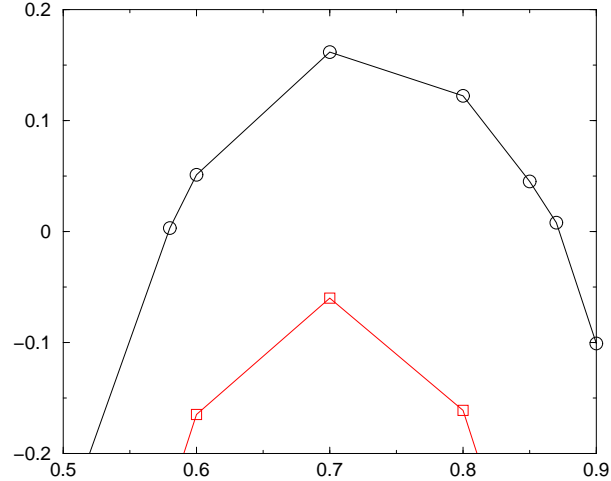}}
\caption{\label{figure:pi2-5_detail} Detail from
Fig. \ref{figure:pi2-5} showing ${\rm Re}\lambda$ for $l=2$ and
$l=3$. The $l=2$ leading mode is unstable for
$0.58 <\kappa< 0.87$.
}
\end{figure}

We have used GD1, GD2 and CD3. All three codes have numerical modes
that are not physical. Often it is clear that they are related to weak
modes at the light cone, and they can be discarded by inspection.
Weak modes are particularly easy to spot in GD1, as it is the most
causal code. In the other codes, the numerical counterparts of weak
modes can appear quite smooth. In GD1 we can suppress weak modes
completely by rearranging the grid spacing so that the light cone
falls exactly on a grid point, and then truncating the grid at that
point. This code will be referred to as GD1LC. In the other codes we
need to pick out the physical modes by the criterion of convergence
both with resolution for a single code and between different
codes. Fig. \ref{figure:pi2_conv} demonstrates convergence of
$\lambda$ for the case $\kappa=1/3$ and $l=2$. The best values (either
from GD2 or CD3) for the Lyapunov exponents are plotted as a function
of $\kappa$ and $l$ in Fig. \ref{figure:pi2-5}. The $l=2$ leading mode
is unstable for $0.58 \lesssim\kappa\lesssim 0.87$ (see also
Fig. \ref{figure:pi2-5_detail}).

Convergence tests fail to identify any mode as physical at the highest
available resolution for $\kappa=0.05$ at $l=4$ and $\kappa=0.1$ at
$l=5$. However, all modes of all three codes decay for these values of
$\kappa$ and $l$, so that we are fairly certain that there are no
physical growing modes for these values of $\kappa$ and
$l$. Furthermore, for all $\kappa$, ${\rm Re}\lambda$ decreases with
increasing $l$ (see Fig. \ref{figure:pi2-5}). Therefore we conclude
that all $l\ge 3$ modes are stable.

\subsection{Polar $l=1$ perturbations} 
\label{section:polar1}

In this section, we discuss the equations of motion for the $l=1$
(dipole) polar perturbations.  There are no dipole gravitational
waves, and therefore the gravitational field has no degrees of freedom
independently of the matter. There are three matter degrees of
freedom. They are an azimuthal fluid velocity perturbation $\alpha$, a
radial fluid velocity perturbation $\gamma$, and a density
perturbation $\omega$. (For the equation of state we use here, the
pressure and density perturbations are related by $\delta
p=\kappa\delta \rho$). The metric perturbations are obtained from the
matter perturbations by constraints. For a general discussion we refer
the reader to \cite{fluidpert}. Here we only carry out the reduction
of the equations given there to a self-similar background solution.

We use the matter variables 
\begin{equation}
\overcirc\alpha\equiv r^{-1}\alpha, \qquad
\overcirc\gamma\equiv -(1+\kappa)\gamma, \qquad
\overcirc\omega\equiv \omega. 
\end{equation}
The matter perturbations are regular at the center if
$\overcirc\alpha$ and $\overcirc\gamma$ are $O(1)$ and even in powers
of $x$, and $\overcirc\omega$ is $O(x)$ and odd. The leading orders of
$\alpha$ and $\gamma$ are additionally constrained as
$(1+\kappa)\overcirc\alpha+\overcirc\gamma=O(x^2)$.

The perturbation variables are not completely gauge-invariant for
$l=1$. We fix most of the gauge freedom by setting the metric
perturbation $k=0$. In this (partially fixed) gauge the remaining
metric perturbations $\psi$, $\chi$ and $\eta$ are determined by
constraints. We introduce suitably rescaled metric perturbation
variables
\begin{equation}
\overcirc\chi\equiv e^{-3\tau} r^{-3}\chi, \qquad
\overcirc\psi= e^{-2\tau} r^{-2}\psi, \qquad
\overcirc\eta=e^{-\tau} r^{-1}\eta. 
\end{equation}
The metric perturbations are regular at the center if $\overcirc\chi$,
$\overcirc\eta$ and $\overcirc\psi$ are all $O(1)$ and even in
$x$. The perturbed spacetime is CSS if and only if all perturbations
are independent of $\tau$. 

From Eqs. (99,100) of \cite{fluidpert}, the equations of motion of the
matter perturbations are
\begin{eqnarray}
\overcirc\gamma\,\dot{}-\kappa\overcirc\omega' & = & - \bar S_\gamma \\
\overcirc\omega\,\dot{}-\overcirc\gamma' & = & -\bar S_\omega \\
\overcirc\alpha\,\dot{} & = & -r^{-1} S_\alpha 
- U\overcirc\alpha ,
\end{eqnarray}
where the expressions $\bar S_\gamma$, $\bar S_\omega$ and $S_\alpha$
are given by Eqs. (A9-11) of \cite{fluidpert}. The first two equations
constitute a first-order form of a wave equation whose characteristics
have velocity $\sqrt{\kappa}$ with respect to the background fluids:
they describe sound waves. The third equation transports
$\overcirc\alpha$ along the fluid. Solving the three matter equations
for the $\tau$-derivatives, on a continuously self-similar background
spacetime, one obtains the following system of evolution equations:
\begin{eqnarray}
\label{l=1_evolution}
\overcirc\gamma_{,\tau} & = & A_\kappa \overcirc\gamma_{,x} 
+ B_\kappa \overcirc\omega_{,x} 
+ S(\overcirc\gamma), \\
\overcirc\omega_{,\tau} & = & C_\kappa \overcirc\gamma_{,x} 
+ A_\kappa \overcirc\omega_{,x} 
+ S(\overcirc\omega), \\
\overcirc\alpha_{,\tau} & = & F \overcirc\alpha_{,x} 
+ S(\overcirc\alpha), 
\end{eqnarray}
The coefficients of the $3\times 3$ matrix $A$ in
(\ref{evolution_equations}) are
\begin{equation}
\label{soundcoeffs}
A_\kappa = -x-{(1-\kappa)V\over(1-\kappa V^2)sg}, 
\quad B_\kappa = {\kappa(1-V^2)\over (1-\kappa V^2)sg}, 
\quad C_\kappa = {(1-V^2)\over (1-\kappa V^2)sg}, 
\quad F= -x-{V\over sg}. 
\end{equation}
Its eigenvalues, the characteristic speeds, are
\begin{equation}
\lambda_{\kappa\pm}=-x-{(1-\kappa)V\over(1-\kappa V^2)sg}
\pm {\sqrt{\kappa}(1-V^2)\over (1-\kappa V^2)sg}, \qquad
\lambda_0 =  -x-{V\over sg}.
\end{equation}
The constraint equations for the metric perturbations are also given
in \cite{fluidpert}. The derivative operator $D$ becomes a simple
partial derivative in coordinates $x$ and $\tau$, namely
$rD=x\partial/\partial x$. We obtain a system of three ODEs in $x$,
from now on referred to as the constraints:
\begin{equation}
\label{l=1_constraints}
xu_{,x}=Mu+s, \qquad u=(\overcirc\chi,\overcirc\eta,\overcirc\psi).
\end{equation}
$M$ and $s$ are given in Appendix \ref{appendix:l=1polarequations}.

We now return to the issue of the gauge freedom that is left after one
has set $k=0$.  The change of $k$ under an arbitrary gauge
transformation parameterized by the variable $\overcirc\xi$ is
$\overcirc k\to \overcirc k+L_1 \overcirc\xi$ (see \cite{fluidpert}
for details.)  $L_1$ is a differential operator (roughly speaking a
second $x$-derivative), whose kernel is parameterized by one free
function of $\tau$. The remaining gauge freedom can therefore be fixed
by imposing a gauge condition worth one function of $\tau$. There are
two natural choices. In ``$\eta$ gauge'' we set $\overcirc\eta=O(x^2)$
at the center. In ``$\alpha$ gauge'' we set $\overcirc\alpha=O(x^2)$,
and therefore also $\overcirc\gamma=O(x^2)$. Numerically it is not
easy to enforce that any variable behaves as $O(x^2)$ at the center.
Therefore we impose the two sub-gauges using variables rescaled by
suitable powers of $x$. This is discussed Appendix
\ref{appendix:l=1polarequations}. 

Although by function counting either $\alpha$ gauge or $\eta$ gauge
should fix the residual gauge freedom in $k=0$, a single gauge mode
does in fact survive in each gauge.  The change of $\overcirc \eta$
under a gauge transformation is $\overcirc\eta\to\overcirc\eta+L_2
\overcirc\xi$. Any residual $\overcirc\xi$ must now obey both
$L_1\overcirc\xi=0$, from the gauge condition $k=0$, and
$L_2\overcirc\xi=O(x^2)$, from the gauge condition
$\overcirc\eta=O(x^2)$, where $L_2$ is another differential operator
(roughly speaking the wave operator). A careful calculation shows that
the two joint equations still have one non-zero solution, which is
$\overcirc\xi_0=e^{\lambda \tau}f(x)$ for
\begin{equation}
\label{gaugemode}
\lambda={1\over2}\pm i\sqrt{\left(\kappa+{1\over
3}\right)\bar\rho(0)-{1\over 4}}.
\end{equation}
That means that in the $\eta=O(x^2)$ sub-gauge of the $k=0$ gauge we
have a single complex conjugate pair of gauge modes left. 

\begin{figure}
\epsfxsize=8cm \centerline{\epsffile{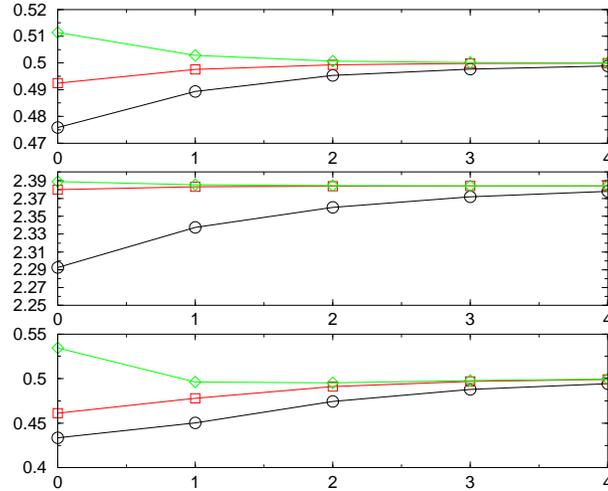}}
\caption{\label{figure:even1_gauge_conv} Convergence of $\lambda$ for
the single gauge mode in the $l=1$ polar perturbations, for
$\kappa=1/3$. From left to right $N=10,20,\dots 160$. Circles are GD1,
squares are GD2, and diamonds CD4. The top two graphs were obtained in
$\eta$ gauge, using the variables of Appendix A, and show ${\rm
Re}\lambda$ and ${\rm Im}\lambda$ for the gauge mode. The value of
$\lambda$ for this mode computed from the background solution and
Eq. (\ref{gaugemode}) is $\lambda\simeq 1/2+2.384i$ for
$\kappa=1/3$. The bottom graph was obtained in $\alpha$ gauge, using
the variables described in appendix B. The exact value of $\lambda$
for this gauge mode is $1/2$ for $\kappa=1/3$. }
\end{figure}

The situation is similar in $\alpha$ gauge: The change of
$\overcirc\alpha$ under a gauge transformation is
$\overcirc\alpha\to\overcirc\alpha+L_3 \overcirc\xi$ for some operator
$L_3$, roughly speaking a first $\tau$-derivative. The joint kernel of
$L_1\overcirc\xi=0$ and $L_3\overcirc\xi=O(x^2)$ is not empty either
but contains a single real gauge mode with
\begin{equation}
\lambda={1+3k\over3(1+k)}.
\end{equation}
Both gauge modes are in fact found numerically, and give us a check
on the numerical precision for each value of
$\kappa$. Fig. \ref{figure:even1_gauge_conv} demonstrates this for
both gauges. 

\begin{figure}
\epsfxsize=8cm \centerline{\epsffile{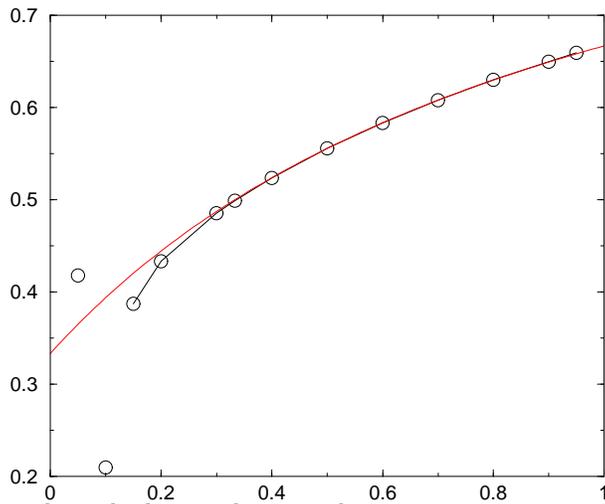}}
\caption{\label{figure:even1_gauge} $\lambda$ for the single gauge
mode in the $l=1$ polar perturbations in $\alpha$ gauge as a function
of $\kappa$. Circles are measured values, connected by straight lines. The
smooth curve is the exact value. The points at $\kappa=0.1$ and $\kappa=0.05$
represent the real part of the only growing mode present, which is
complex.}
\end{figure}

\begin{figure}
\epsfxsize=8cm \centerline{\epsffile{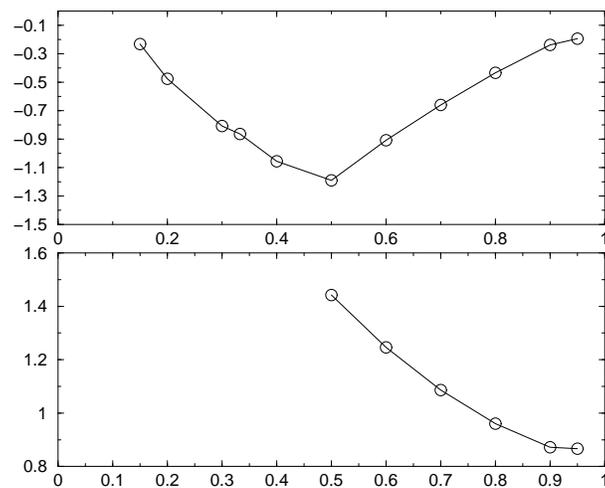}}
\caption{
\label{figure:even1} 
Best value (obtained with GD2) for the physical Lyapunov exponent
$\lambda$ of the $l=1$ polar perturbations. The upper graph shows
${\rm Re}\lambda$ against $\kappa$, and the lower graph ${\rm
Im}\lambda$. Points are connected by straight lines. The measured
values end at $\kappa=0.15$ because for $\kappa=0.1$ and $\kappa=0.05$
the top mode could not be identified. The graph of ${\rm Im}\lambda$
ends because $\lambda$ is real for $\kappa< 0.5$. Clearly a real mode
and a complex mode pair have crossed between $\kappa=0.5$ and
$\kappa=0.6$.  }
\end{figure}

We have used GD1, GD2 and CD4.  For $\kappa\gtrsim 0.15$, the numerical
results are straightforward. Both in $\alpha$ gauge and in $\eta$
gauge, we find the expected gauge mode. We find that $\alpha$ gauge,
in the specialized variables described in Appendix
\ref{appendix:l=1polarequations}, has the best numerical behavior at
the center. As an estimate of numerical error in the code,
Fig. \ref{figure:even1_gauge} shows the exact and numerical value of
$\lambda$ for the gauge mode in $\alpha$
gauge. Fig. \ref{figure:even1} shows the best value of $\lambda$ in
this system as a function of $\kappa$.

For $\kappa\lesssim 0.15$, our code does not work well.
Fig. \ref{figure:even1_gauge} shows that with decreasing $\kappa$, the
error in locating the gauge mode increases rapidly. For $\kappa=0.1$
and $\kappa=0.05$, all three codes fail to find the real gauge
mode. Nevertheless, GD1 and GD2 have precisely one growing mode, and
CD4 has precisely one growing mode that is not clearly a grid
mode. All these modes are complex, but they obey $|{\rm
Re}\,u(x)|\gg|{\rm Im}\,u(x)|$ for all $x$, by a factor of $\sim100$:
the mode is almost real, modulated by a $\cos\omega\tau$ factor. It is
also clear that there is a finite differencing problem at the center
that affects all three codes: the functions $u(x)$ are not
well-behaved even or odd functions. The origin of this instability is
uncertain. A plausible explanation is that the numerical background
solution itself is not sufficiently well-behaved at the center for
small $\kappa$. Some of the coefficients in the $l=1$ (and also $l\ge
2$) polar perturbation equations are very large and sharply peaked at
the center. Furthermore, some of these coefficients must be obtained
as numerical derivatives of the background solution. These numerical
derivatives are not smooth at the center, and we had to artificially
extrapolate them to the center in order to smoothe them.

Because all three codes agree on a single growing mode that is almost
real in the sense just defined, we identify this mode with the gauge
mode, even though the agreement in $\lambda$ is poor. This leaves no
other growing modes in any of the codes for any $\kappa$ (except for
obvious grid modes in CD4 that can be ruled out by inspection.) We
therefore conjecture that there are no growing physical modes, even
though we cannot identify the top physical mode.

\subsection{Polar $l\ge 2$ perturbations} 
\label{section:polarge2}

As discussed in detail in \cite{fluidpert}, the $l\ge2$ polar
perturbations of a spherical perfect fluid admit a free evolution
formulation. The dynamical variables that can be specified freely are
the metric perturbations $\chi$, $k$ and $\psi$, and the time
derivatives $\dot\chi$ and $\dot k$. $\chi$ and $k$ obey wave
equations, and $\psi$ a transport equation, all of which are coupled.
There are three gauge-invariant matter perturbations. $\alpha$ is an
azimuthal velocity perturbation, $\gamma$ is a radial velocity
perturbation, and $\omega$ is a pressure and density
perturbation. These matter perturbation variables are completely
determined as algebraic expressions in the seven first-order metric
perturbation variables, as discussed in detail in
\cite{fluidpert}. For the purpose of finding the late-time behavior of
generic perturbations, we work with the metric perturbations alone.

Solving for the $\tau$-derivative of the seven variables, we have find
that the matrix $A$ in (\ref{evolution_equations}) takes the following
form:
\begin{equation}
A=\left(\begin{array}{cccccccc}
A_1 & B_1 & & & D \\
C_1 & A_1 & & & vD \\
& & A_\kappa & B_\kappa & E \\
& & C_\kappa & A_\kappa & vE \\
& & & & F \\ 
\end{array}\right),
\qquad u=
\left(\begin{array}{c}
\tilde\chi \\ \hat\chi \\ \tilde k \\ \hat k \\ \overcirc\psi \\
\end{array}\right).
\end{equation}
The remaining two variables $\overcirc\chi$ and $\overcirc k$ are
evolved using only source terms. The coefficients of the matrix $A$
have already been given in Eqs. (\ref{lightcoeffs}) and
(\ref{soundcoeffs}), except for
\begin{equation}
D = {2(\bar\mu-\bar U)\over s^2gx}, 
\quad E= -{2\kappa x\bar U(1-V^2)\over(1-\kappa V^2)g}.
\end{equation}
The five eigenvalues of $A$ are $\lambda_{1\pm}$,
$\lambda_{\kappa\pm}$ and $\lambda_0$, which have already been
given. ($D$ and $E$ do not influence the eigenvalues.)  Note that $s$
has been chosen so that $\lambda_{\kappa+}>0$ for $0\le x<1$. The
point $x=1$ where it changes sign is the sound cone. Similarly, the
point $x=x_{\rm lc}>1$ where $\lambda_{1+}$ changes sign is the light
cone. Clearly $x_{\rm lc}$ is 1 for $\kappa=1$ and diverges as
$\kappa\to0$. Note that the $(\tilde\chi,\hat\chi)$ sub-matrix of $A$
is the same as the entire matrix $A$ for the axial $\Pi$
perturbations, as both pairs of variables obey a wave equation at the
speed of light. Similarly, the $(\tilde k,\hat k)$ sub-matrix is equal
to the entire matrix for the $l=1$ polar perturbations, which obey a
wave equation at the speed of sound. Therefore $A_\kappa$ is equal to
$A_1$ for $\kappa=1$, when the speed of sound is the speed of light.

\begin{figure}
\epsfxsize=8cm
\centerline{\epsffile{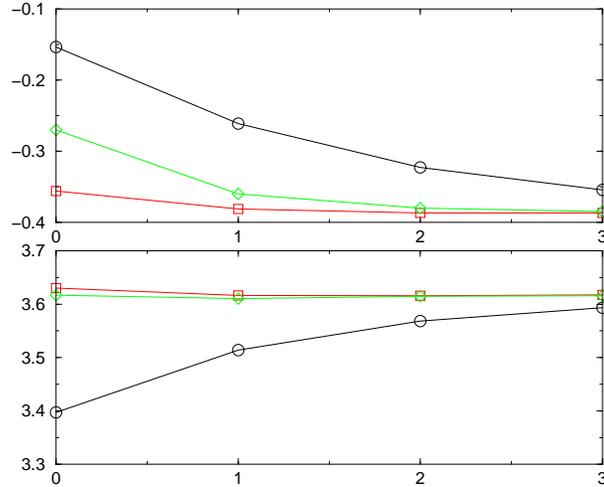}}
\caption{\label{figure:even2_k=0.333_conv} Convergence of $\lambda$ for the
top $l=2$ polar mode with resolution at $\kappa=1/3$. From left to right
$N=20,40,\dots 160$. Circles are GD1, squares are GD2, and diamonds
CD3. The upper
graph is ${\rm Re}\lambda$, and the lower graph is ${\rm
Im}\lambda$.}
\end{figure}

\begin{figure}
\epsfxsize=8cm \centerline{\epsffile{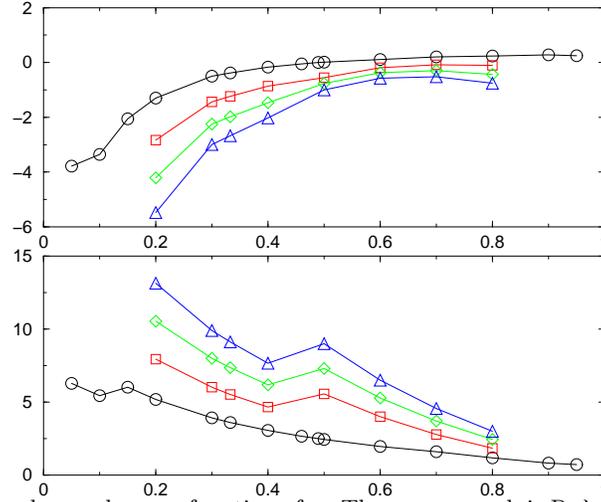}}
\caption{\label{figure:even2-5} $\lambda$ for the top $l=2...5$ polar
modes as a function of $\kappa$. The upper graph is ${\rm Re}\lambda$, and
the lower graph is ${\rm Im}\lambda$. Circles are $l=2$, squares
$l=3$, diamonds $l=4$ and triangles $l=5$. Note that ${\rm Re}\lambda$
decreases with $l$, while ${\rm Im}\lambda$ increases. The curves for
$l=3,4,5$ do not extend to the lowest and highest $\kappa$ because of large
numerical error.}
\end{figure}

We have used the three codes GD1, GD2 and CD3. There is good agreement
and convergence between the three codes for $0.2\lesssim \kappa
\lesssim 0.8$. Fig. \ref{figure:even2_k=0.333_conv} demonstrates
convergence of $\lambda$ for $\kappa=1/3$ and
$l=2$. Fig. \ref{figure:even2-5} shows the best values of $\lambda$
for $l=2\dots 5$.  We find that all modes for all $l$ decay for
$\kappa\lesssim 0.49$. For $\kappa\gtrsim 0.49$, there is an unstable
$l=2$ mode. At still higher $\kappa$, there is probably more than one
unstable mode, but it is difficult to identify subdominant modes with
certainty.

\begin{figure}
\epsfxsize=8cm
\centerline{\epsffile{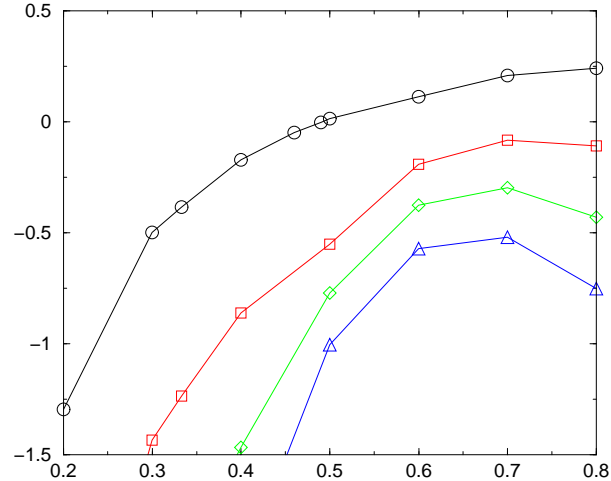}}
\caption{\label{figure:even2-5_detail} 
Detail of Fig. \ref{figure:even2-5}. ${\rm Re}\lambda$ is positive for
$l=2$, $\kappa> 0.49$. ${\rm Re}\lambda$ is negative for all $\kappa$ for
$l>2$. }
\end{figure}

For $\kappa\gtrsim 0.8$, and $\kappa\lesssim0.2$, the numerical
precision decreases quickly. In this region the numerical accuracy is
poor for $l=2$, and for $l=3,4,5$ even the top physical mode cannot be
identified clearly. The origin of these problems is different for high
and low $\kappa$. At high $\kappa$, all modes in all three codes show
discontinuities at the sound cone. These are not due to the weak modes
discussed above, but are due to unsmooth behavior in the numerical
background solution, through certain coefficients of the perturbation
equations that must be obtained as numerical derivatives of the
background solution. At low $\kappa$, the origin of the low precision
is not so clear. It may be that the background coefficients become
increasingly large and sharply peaked in $x$ near the center as
$\kappa\to0$. With the sound cone at $x=1$, the value of $x$ at the
light cone also diverges as $1/\sqrt{k}$ as the sound speed goes to
zero. (Recall that $x$ is defined so that the sound cone is at $x=1$.)
In a second source of error, some of the perturbation coefficients, as
determined from the numerical background solution, must be smoothed at
the center for small $\kappa$, and this introduces additional
error. (This is the same problem that affects the $l=1$ polar modes.)

\begin{figure}
\epsfxsize=8cm
\centerline{\epsffile{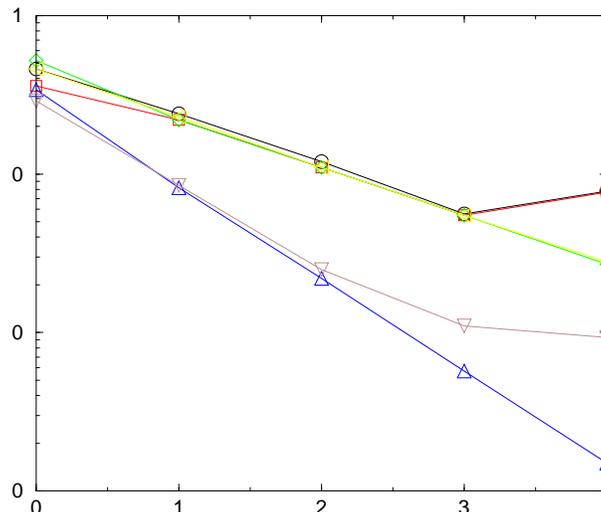}}
\caption{\label{figure:even2_k=0.6_conv} Log-log plot of
independent residual evaluation between three different codes for the
dominant polar $l=2$ mode for $\kappa=0.6$. Evaluating the mode produced by
GD2 with CD3 shows exact second-order convergence (triangles up, bottom
curve). The opposite operation, checking on CD3 with GD2, shows
second-order convergence that breaks down at high resolution
(triangles down). Checking on CD3 and GD2 with GD1 shows the expected
first-order convergence (triangles left and diamonds). The opposite
check shows first-order convergence that breaks down at low
resolution (squares and circles, top curves). The resolutions shown
are $\Delta x=1/10\dots 1/160$.}
\end{figure}

In contrast to the inadequacy of the numerics at high and low
$\kappa$, the numerical precision is good in a large neighborhood of
the value $\kappa\simeq 0.49$. It is therefore certain that the $l=2$
polar perturbations are stable for $\kappa\lesssim 0.49$ and unstable
for $\kappa\gtrsim 0.49$. As evidence that the instability is
physical, Fig. \ref{figure:even2_k=0.6_conv} demonstrates the
convergence of independent residual evaluations between all three
codes for $l=2$ and $\kappa=0.6$. From convergence, we estimate
$\lambda$ (for this $\kappa$ and $l$) as $\lambda=(0.112\pm0.003)
+(1.968\pm0.005) i$. The real part of $\lambda$ is therefore much
larger than the finite differencing error.

\section{Conclusions}
\label{section:results}

The threshold of gravitational collapse of a perfect fluid with
$p=k\rho$ in spherical symmetry shows ``type II'' critical phenomena
(including mass scaling with a critical exponent) for the entire range
$0<k<1$ \cite{EvansColeman,Maison,KHA2,NeilsenChoptuik}. Is this true
also when one relaxes the assumption of spherical symmetry? We have
addressed this question by studying the nonspherical perturbations of
the critical solutions in spherical symmetry. We have extended the
results of \cite{codim} from $\kappa=1/3$ in the equation of state
$p=k\rho$ to the entire physical range $0<k<1$. Because of the
numerical difficulties, we have complemented the single first-order
scheme used in \cite{codim} with three different second-order finite
differencing schemes. We have complemented convergence tests by
independent residual evaluation. This allows us to identify ``weak
modes'', numerical artifacts related to weak solutions of the
perturbation equations. Finally, we have corrected errors in
\cite{codim} concerning the equations of motion for the $l=1$ axial
and polar perturbations.

After verifying that the CSS solution has exactly one growing
spherical perturbation for all $\kappa$, we have found that all
nonspherical perturbations decay for all $\kappa$ with the following
exceptions:

1. For $\kappa<1/9$, there is precisely one growing $l=1$ axial
mode. This result was obtained analytically, even though the
background solution is known only numerically. Because it is $l=1$,
this mode is three-fold degenerate. 

2. For $\kappa\gtrsim 0.49$, there is a growing $l=2$ polar mode. We
cannot rule out that at larger $\kappa$, there are several growing
modes. Because it is $l=2$, this mode is five-fold degenerate.

3. For $0.58\lesssim \kappa\lesssim 0.87$, there is also a growing
$l=2$ axial mode.

The numerical evidence for this is not as good as one would
like. Because of numerical error, we cannot measure $\lambda$ of the
top physical mode of the $l=1$ polar perturbations for $\kappa\lesssim
0.15$, nor of the $l\ge 3$ polar perturbations for $\kappa\lesssim
0.2$ and $\kappa\gtrsim 0.8$. However, in all these cases where
physical modes and numerical modes cannot be distinguished clearly,
all modes, including the physical modes, do in fact decay. Therefore
we still argue that all physical modes decay. One cause of the
numerical difficulty is the separate existence of light cones and
sound cones, which gives rise to numerical artifacts related to weak
solutions of the continuum equations. A second cause is that instead
of a single smooth background solution we are dealing with a
one-parameter family of such solutions, which is ill-behaved at both
ends $\kappa=0$ and $\kappa=1$. A third cause is that several
coefficients required in the perturbation equations are first and
second derivatives of the background fields, which are not perfectly
smooth at the center and the light cone.

What is the significance of our results? The $l=1$ axial perturbations
are naturally associated with infinitesimal (differential)
rotation. The presence of a growing rotation mode at low $\kappa$
($<1/9$) is not surprising, as one would expect a rotating dust
configuration to be torn apart by centrifugal forces. For a
sufficiently stiff fluid this intuition obviously fails. The
significance of the $l=1$ axial perturbations is that they can survive
into the final black hole formed in collapse (and turn a Schwarzschild
black hole into a Kerr black hole), while all other non-spherical
perturbations must be radiated during collapse. Their instability is
expected to give rise to interesting new phenomena in critical
collapse that will be explored elsewhere \cite{scalingfunctions}.

The existence of a small number of unstable polar modes at high $\kappa$
($\gtrsim 0.49$) is more puzzling. At face value it suggests that the
spherically symmetric CSS solution is a critical solution only when
restricted to exact spherical symmetry. Its place may be taken by
another critical solution with less symmetry, or there may not be a
critical solution, and hence no universality, at the black hole
threshold.

\acknowledgments

I am grateful to Jos\'e M. Mart\'\i n-Garc\'\i a for pointing
out that I was using the wrong equations of motion for $l=1$
perturbations in \cite{codim}, and for help in correcting them. I
would like to thank Bob Wald for helpful conversations on the $l=1$
axial perturbation equation, and Miguel Alcubierre, Matt Choptuik,
Giampaolo D'Alessandro and David Garfinkle for suggestions on
numerical methods. This research was supported in part by NSF grant
PHY-95-14726 to the University of Chicago, and by EPSRC grant
GR/N10172.

\appendix

\section{Background equations}
\label{appendix:bgeqns}

The relation between the fluid frame derivatives and the partial
derivatives in CSS coordinates is
\begin{eqnarray}
\label{defdot}
\dot f & = & {e^\tau g\over a\sqrt{1-V^2}}
\left[f_{,\tau}+\left(x+{V\over sg}\right)f_{,x}\right], \\
\label{defprime}
f' & = & {e^\tau g\over a\sqrt{1-V^2}}
\left[Vf_{,\tau}+\left(Vx+{1\over sg}\right)f_{,x}\right], \\
rDf & = & x f_{,x}.
\end{eqnarray}
The perturbation equations of \cite{fluidpert} allow for a 2-parameter
equation of state $p=p(\rho,s)$, where $s$ is the entropy per
particle. When restricting to our simple barotropic equation of state,
we set both the fluid entropy $s$ and its gauge-invariant perturbation
$\sigma$ to zero. We also set $C\equiv\partial p/\partial s$ to zero,
and we set the sound speed squared $c_s^2\equiv\partial p/\partial
\rho$ equal to the constant $\kappa$.

The background equations that result from the CSS ansatz are
\begin{equation}
\label{aconstraint}
a^{-2} = 1-{2s^2x^2\bar\rho \over 1-V^2}
\left[1+\kappa V^2+{(1+\kappa)V\over sgx}\right]
\end{equation}
\begin{eqnarray}
\label{rhox}
{d\ln\bar\rho\over dx} & = &(1+\kappa)D^{-1}[(V+gsx)S_1-(1+Vgsx)S_2], \\
\label{vx}
{dV\over dx} & = & (1-V^2)D^{-1}[-\kappa(1+Vgsx)S_1+(V+gsx)S_2], \\
\label{gx}
x{d\ln g\over dx} & = & 1-a^2+(1-\kappa)x^2s^2a^2\bar\rho,
\end{eqnarray}
where 
\begin{eqnarray}
S_1 & \equiv & 2{V\over x} + {2gs\over 1+\kappa} + VC_\alpha + C_a , \\
S_2 & \equiv &{2\kappa Vgs\over 1+\kappa} + C_\alpha + VC_a, \\
D & \equiv & \kappa(1+Vgsx)^2 - (V+gsx)^2, \\
C_a & \equiv & {ra_{,t}\over x\alpha} =
-xs^2\bar\rho a^2(1+\kappa){V\over 1-V^2} = O(x^2), \\
C_\alpha & \equiv & {r\alpha_{,r}\over x\alpha} = 
{a^2-1\over 2x} + s^2x\bar\rho a^2\left({1+\kappa\over1-V^2}-1\right)
= O(x).
\end{eqnarray}
Here (\ref{rhox}) and (\ref{vx}) are the fluid equations of motion,
and (\ref{gx}) and (\ref{aconstraint}) are two of the Einstein
equations. The terms $C_a$ and $C_\alpha$ arise when metric
derivatives in the matter equations are eliminated using the Einstein
equations. The expression for $a$ incorporates the regularity
condition (absence of a conical singularity) $a=1$ at $r=0$.

The following background quantities are required as coefficients in
the perturbation equations.
\begin{eqnarray}
\bar W & \equiv & e^{-\tau}W\equiv e^{-\tau} n^Ar_{,A}  
= (asx)^{-1}(1-V^2)^{-1/2}=O(x^{-1}), \qquad
\bar U \equiv e^{-\tau}U\equiv e^{-\tau} u^Ar_{,A} = V\bar W = O(1), \\
\bar\mu & \equiv & e^{-\tau}\mu \equiv e^{-\tau} {u^A}_{|A} =O(1), \qquad
\bar\nu  \equiv  e^{-\tau}\nu \equiv e^{-\tau} {n^A}_{|A} =O(x), \\
\bar\rho & \equiv & e^{-2\tau} 4\pi\rho = O(1), \qquad
{2m\over r} \equiv  1-r_{,A}r^{,A} = 1- r^2(W^2 - U^2) = 1-a^{-2}= O(x^2).
\end{eqnarray}
In a CSS background all these quantities depend only on $x$, and we
have indicated their behavior at the origin.

\section{$l=1$ polar perturbation equations} 
\label{appendix:l=1polarequations}

The source terms in Eq. (\ref{l=1_evolution}) are obtained from the
equations of \cite{fluidpert} in the form 
\begin{eqnarray}
S(\overcirc\gamma)&=&-e^{-\tau}{a\sqrt{1-V^2}\over g(1-\kappa V^2)} 
(\bar S_\gamma+\kappa V \bar S_\omega), \\
S(\overcirc\omega)&=&-e^{-\tau}{a\sqrt{1-V^2}\over g(1-\kappa V^2)}
(V\bar S_\gamma+\bar S_\omega), \\
S(\overcirc\alpha)&=&-e^{-\tau}{a\sqrt{1-V^2}\over g} \left(r^{-1} S_\alpha 
+ U\overcirc\alpha\right) ,
\end{eqnarray}
where the terms on the right-hand side are defined in
Ref. \cite{fluidpert}. Fully expanded in the variables adapted to
self-similarity, they are
\begin{eqnarray}
\label{Sgamma}
S(\overcirc\gamma)&=&{a\sqrt{1-V^2}\over g(1-\kappa V^2)}\Bigg\{
[(\kappa-1)\bar\mu+4\kappa\bar U+(\kappa-1)V\bar\nu]\overcirc\gamma 
-2\kappa(1+\kappa)V\left[-{1\over sx}+(1+\kappa)sx\bar\rho\right]
\overcirc\alpha, \nonumber \\
&&+{1\over2}(1+\kappa)^2(\bar\mu+V\bar\nu)(sx)^2\overcirc\psi 
-(1+\kappa)[\bar\nu-\bar W+\kappa V(\bar\mu+U)](sx)\overcirc\eta
+(1+\kappa)(\bar\nu+\kappa V\bar\mu)(sx)^3\overcirc\chi\Bigg\}, \\
S(\overcirc\omega)&=&{a\sqrt{1-V^2}\over g(1-\kappa V^2)} \Bigg\{
\left[(\kappa-1)V\bar\mu+2\kappa V\bar U
+{\kappa-1\over\kappa}\bar\nu+2\bar W\right]\overcirc\gamma 
-2(1+\kappa)\left[-{1\over sx}+(1+\kappa)sx\bar\rho\right]
\overcirc\alpha \nonumber \\ 
&&+{1\over2}(1+\kappa)\left({1+\kappa\over \kappa}\bar\nu-2\bar W
+(1+\kappa)V\bar\mu+2\kappa V\bar U\right)(sx)^2\overcirc\psi
\nonumber\\
&&-(1+\kappa)(\bar\mu+V\bar\nu)(sx)\overcirc\eta
+(1+\kappa)(\bar\mu+V\bar\nu)(sx)^3\overcirc\chi \Bigg\}, \\
\label{Salpha}
S(\overcirc\alpha)&=&{a\sqrt{1-V^2}\over g}\Bigg\{
-{\kappa\over 1+\kappa}(sx)^{-1}\overcirc\omega +
\left[\kappa\bar\mu+(2\kappa-1)\bar U\right]\overcirc\alpha
-\overcirc\eta +{1\over2}(sx)^2\overcirc\chi
\Bigg\} .
\end{eqnarray}

The coefficients in the constraint equations
Eq. (\ref{l=1_constraints}) are
\begin{eqnarray}
M&=&\left(\begin{array}{ccc}
-1+a^2\left[-2+(1-\kappa)(sx)^2\bar\rho\right] &
a^2(4\bar U\bar W+2\bar\mu \bar W-2\bar\nu \bar U)(sx) &
2a^2(\bar\nu\bar U-\bar\mu \bar W)(sx)^3 \\
-2a^2\bar U\bar W(sx)^3 &
2-2a^2[1 -\bar U^2(sx)^2+\kappa\bar\rho(sx)^2] &
-a^2\left[\bar W^2+\bar U^2-(1+\kappa)\bar\rho\right](sx)^4 \\
2a^2\left[\bar U\bar\nu-(\bar\mu+\bar U)\bar W\right](sx) &
-2a^2\bar U^2 &
-3-2a^2\left[1+\bar U^2(sx)^2-(sx)^2\bar\rho\right] \\
\end{array}\right) \\
s&=&a^2\Big(-2\bar\rho\overcirc\gamma 
+4\bar\rho(1+\kappa)\bar W(sx)\overcirc\alpha, \quad
(1-\kappa)\bar\rho(sx)\overcirc\omega -
4(1+\kappa)\bar\rho \bar U (sx)^2\overcirc\alpha, \quad
2\bar\rho(sx)^{-1}\overcirc\omega -
8(1+\kappa)\bar\rho \bar U\overcirc\alpha\Big).
\end{eqnarray}

In ``$\eta$ gauge'' we set $\overcirc\eta=O(x^2)$ at the center.  The
constraint equations are then solved with the boundary conditions
\begin{equation}
\label{etagaugeBC}
\overcirc\psi=2(1+\kappa)\bar\rho\overcirc\alpha, \qquad
\overcirc\eta=0, \qquad
\overcirc\chi={2\over 5}\bar\rho(sx)^{-1}\overcirc\omega
+{32\over 15}\bar\rho\overcirc\alpha
\end{equation}
at the center $x=0$.  The condition $\overcirc\eta=0$ is our 
gauge condition, and the other two conditions follow from $Mu+s=0$
($M$ has rank 2 at the center). 
An alternative way of imposing $\eta$ gauge is to introduce the new
variable
\begin{equation}
\tilde\eta=(sx)^{-2}\overcirc\eta,
\end{equation}
which is $O(1)$ at the center in this gauge. The evolution equations
are as before, with only $\overcirc\eta$ replaced by
$(sx)^2\tilde\eta$ in (\ref{Sgamma}-\ref{Salpha}). The constraint
equations become
\begin{equation}
x\tilde u_{,x}=\tilde M\tilde u+\tilde s, \qquad
\tilde u=(\overcirc\chi,\tilde\eta,\overcirc\psi),
\end{equation}
with
\begin{eqnarray}
\tilde M&=&\left(\begin{array}{ccc}
-1+a^2\left[-2+(1-\kappa)(sx)^2\bar\rho\right] &
a^2(4\bar U\bar W+2\bar\mu \bar W-2\bar\nu \bar U)(sx)^3 &
2a^2(\bar\nu\bar U-\bar\mu \bar W)(sx)^3 \\ 
-2a^2\bar U\bar Wsx &
-2a^2[1 -\bar U^2(sx)^2+\kappa\bar\rho(sx)^2] &
-a^2\left[\bar W^2+\bar U^2-(1+\kappa)\bar\rho\right](sx)^2 \\
\big(2a^2\left[\bar U\bar\nu-(\bar\mu+\bar U)\bar W\right](sx) &
-2a^2\bar U^2(sx)^2 &
-3-2a^2\left[1+\bar U^2(sx)^2-(sx)^2\bar\rho\right] \\
\end{array}\right) \\
\tilde s&=&a^2(-2\bar\rho\overcirc\gamma 
+4\bar\rho(1+\kappa)\bar W sx\overcirc\alpha, \quad
(1-\kappa)\bar\rho(sx)^{-1}\overcirc\omega -
4(1+\kappa)\bar\rho \bar U\overcirc\alpha, \quad
2\bar\rho(sx)^{-1}\overcirc\omega -
8(1+\kappa)\bar\rho \bar U\overcirc\alpha).
\end{eqnarray}
The boundary conditions at $x=0$ are
\begin{equation}
\overcirc\psi=2\bar\rho(1+\kappa)\overcirc\alpha, \qquad
\tilde\eta=-{1\over 5}\bar\rho(sx)^{-1}\overcirc\omega
+{4\over 15}\bar\rho\overcirc\alpha, \qquad
\overcirc\chi={2\over 5}\bar\rho(sx)^{-1}\overcirc\omega
+{32\over 15}\bar\rho\overcirc\alpha,
\end{equation}
from $Mu+s=0$ ($M$ has rank 3 at the center). 

In ``$\alpha$ gauge'' we set $\overcirc\alpha=O(x^2)$ (and therefore
also $\overcirc\gamma=O(x^2)$).  The constraints are then solved with
the boundary conditions
\begin{equation}
\label{alphagaugeBC}
\overcirc\psi={4\kappa\over3(1+\kappa)^2}(sx)^{-1}\overcirc\omega,
\qquad
\overcirc\eta=-{\kappa\over1+\kappa}(sx)^{-1}\overcirc\omega, 
\qquad
\overcirc\chi=\left({2\over 5}\bar\rho
+{8\kappa\over9(1+\kappa)^3}\right)(sx)^{-1}\overcirc\omega.
\end{equation}
Here the boundary condition on $\eta$ comes from consistency with the
evolution equation $\overcirc\alpha_{,\tau}=O(x^2)$.
An alternative way of imposing $\alpha$ gauge is to introduce new
matter variables
\begin{equation}
\hat\omega\equiv (sx)^{-1}\overcirc\omega, \qquad
\hat\gamma\equiv \kappa^{-1}(sx)^{-1}\overcirc\gamma, \qquad
\check\alpha\equiv (sx)^{-1} \overcirc\alpha=sx
\hat\alpha.
\end{equation}
The evolution equations
in these variables are
\begin{eqnarray}
\hat\omega_{,\tau} & = & A_\kappa \hat\omega_{,x} 
+ B_\kappa \hat\gamma_{,x} 
+ S(\hat\omega), \\
\hat\gamma_{,\tau} & = & C_\kappa \hat\omega_{,x} 
+ A_\kappa \hat\gamma_{,x} 
+ S(\hat\gamma), \\
\check\alpha_{,\tau} & = & F \check\alpha_{,x} 
+ S(\check\alpha).
\end{eqnarray}
Note that the coefficients of the $x$-derivatives are unchanged, but
that $\hat\omega$ has taken the place of $\overcirc\gamma$ and
$\hat\gamma$ has taken the place of $\overcirc\omega$. Accordingly,
$\hat\gamma$ is now odd and $O(x)$ while $\hat\omega$ is even and
$O(1)$. The source terms are
\begin{eqnarray}
\label{Sgammahat}
S(\hat\gamma)&=&{a\sqrt{1-V^2}\over \kappa g(1-\kappa V^2)}\Bigg\{
[(\kappa-1)\bar\mu+4\kappa\bar U+(\kappa-1)V\bar\nu]\kappa\hat\gamma 
-2\kappa(1+\kappa)V\left[-1+(1+\kappa)(sx)^2\bar\rho\right]
(sx)^{-1}\check\alpha, \nonumber \\
&&+{1\over2}(1+\kappa)^2(\bar\mu+V\bar\nu)sx\overcirc\psi 
-(1+\kappa)[\bar\nu-\bar W+\kappa V(\bar\mu+U)]\overcirc\eta
+(1+\kappa)(\bar\nu+\kappa V\bar\mu)(sx)^2\overcirc\chi\Bigg\}
\nonumber \\
&&-\left[1+{(1-\kappa)V\over(1-\kappa V^2)sgx}\right]\hat\gamma
+{1-V^2\over(1-\kappa V^2) sgx}\hat\omega, \\
S(\hat\omega)&=&{a\sqrt{1-V^2}\over g(1-\kappa V^2)} \Bigg\{
\left[(\kappa-1)V\bar\mu+2\kappa V\bar U
+{\kappa-1\over\kappa}\bar\nu+2\bar W\right]\kappa\hat\gamma 
-2(1+\kappa)\left[-1+(1+\kappa)(sx)^2\bar\rho\right]
(sx)^{-1}\check\alpha \nonumber \\ 
&&+{1\over2}(1+\kappa)\left({1+\kappa\over \kappa}\bar\nu-2\bar W
+(1+\kappa)V\bar\mu+2\kappa V\bar U\right)sx\overcirc\psi
\nonumber\\
&&-(1+\kappa)(\bar\mu+V\bar\nu)\overcirc\eta
+(1+\kappa)(\bar\mu+V\bar\nu)(sx)^2\overcirc\chi \Bigg\}
\nonumber \\
&&-\left[1+{(1-\kappa)V\over(1-\kappa V^2)sgx}\right]\hat\omega
+{\kappa(1-V^2)\over(1-\kappa V^2) sgx}\hat\gamma, \\
\label{Salphacheck}
S(\check\alpha)&=&{a\sqrt{1-V^2}\over g}\Bigg\{
-{\kappa\over 1+\kappa}(sx)^{-1}\hat\omega +
\left[\kappa\bar\mu+(2\kappa-1)\bar U\right]\hat\alpha
-(sx)^{-1}\overcirc\eta +{1\over2}(sx)\overcirc\chi
\Bigg\} 
-\left(1+{V\over sgx}\right)\check\alpha .
\end{eqnarray}
The variable $\check\alpha$, which is odd and $O(x)$, is much better
behaved than the more obvious definition $\hat\alpha\equiv (sx)^{-2}
\overcirc\alpha$. (The combination
$\overcirc\eta+\kappa/(1+\kappa)\hat\omega$ is $O(x^2)$ at the
center. Numerically, it is easier to divide this by $sx$ in
$S(\check\alpha)$ to obtain an $O(x)$ term than to divided by
$(sx)^2$.) The constraints are solved with the matrix $M$ given above
and the source terms
\begin{equation}
\check s=a^2(-2\bar\rho\kappa (sx)\hat\gamma 
+4\bar\rho(1+\kappa)\bar W (sx)^2\check\alpha, \quad
(1-\kappa)\bar\rho(sx)^2\hat\omega -
4(1+\kappa)\bar\rho \bar U (sx)^3\check\alpha, \quad
2\bar\rho\hat\omega -
8(1+\kappa)\bar\rho \bar U(sx)\check\alpha).
\end{equation}
The boundary conditions are (\ref{alphagaugeBC}) with
$(sx)^{-1}\overcirc\omega$ replaced by $\hat\omega$.

\section{$l\ge2$ polar perturbation equations} 
\label{appendix:l>=2polarequations}

In order to work with variables that are regular and $O(1)$ at the
origin for any $l$, we redefine them as $\chi=r^{l+2}\bar\chi$,
$\psi=r^{l+1}\bar\psi$ and $k=r^l\bar k$, as in \cite{fluidpert}. In
order to obtain a first-order formulation, we introduce the frame
derivatives of the barred quantities with respect to the fluid
frame. From Eqs. (87-89) of \cite{fluidpert}, the equations then take
the form
\begin{eqnarray}
(\dot{\bar\chi})\dot{}-(\bar\chi')'-2(\mu-U)r^{-1}\bar\psi'
&=& -r^{-(l+2)}S_\chi -2(l+2)U\dot{\bar\chi}
-(l+2)[(l+2)U^2+\dot U]\bar\chi 
+2(l+2)W\bar\chi' \nonumber \\
&& +(l+2)[(l+2)W^2+W']\bar\chi +2(\mu-U)r^{-1}(l+1)W\bar\psi 
\equiv  e^{(l+4)\tau} S_1 , \\
(\bar\chi')\dot{}-(\dot{\bar\chi})' &=& \nu\dot{\bar\chi}-\mu\bar\chi'
\equiv  e^{(l+4)\tau} S_2, \\
(\dot{\bar k})\dot{}-\kappa(\bar k')'+2\kappa Ur\bar\psi'
&=& -r^{-l}S_k -2lU\dot{\bar k} -l(lU^2+\dot U)\bar k 
+2\kappa lW\bar k'+\kappa l(lW^2+W')\bar k  \nonumber \\
&&
-2\kappa Ur(l+1)W\bar\psi \equiv  e^{(l+2)\tau} S_3 , \\
(\bar k')\dot{}-(\dot{\bar k})' &=& \nu\dot{\bar k}-\mu\bar k'
\equiv  e^{(l+2)\tau} S_4, \\
(\bar\psi)\dot{} &=& 
-r^{-(l+1)}S_\psi-(l+1)U\bar\psi\equiv  e^{(l+2)\tau} S_5,
\end{eqnarray}
where the source terms $S_\chi$, $S_k$ and $S_\psi$ are given in
Eqs. (A1-A3) of \cite{fluidpert}. The second and fourth equations are
just the commutation relations between the dot and prime
derivatives. Here they serve as auxiliary evolution equations for
$\bar\chi'$ and $\bar k'$. To these five equations we add the trivial
evolution equations $(\bar\chi)\dot{}=\dot{\bar\chi}$ and $(\bar
k)\dot{}=\dot{\bar k}$.

For our particular application to a self-similar background we further
rescale these first-order variables to obtain the final dynamical
variables
\begin{eqnarray}
\overcirc\chi & = & e^{-(l+2)\tau} \bar\chi, \qquad
\tilde\chi = e^{-(l+3)\tau} (\bar\chi)\dot{}, \qquad
\hat\chi = e^{-(l+3)\tau} (\bar\chi)', \\
\overcirc k & = & e^{-l\tau} \bar k, \qquad
\tilde k = e^{-(l+1)\tau} (\bar k)\dot{}, \qquad
\hat k = e^{-(l+1)\tau} (\bar k)', \qquad
\overcirc\psi = e^{-(l+1)\tau} \bar\psi,
\end{eqnarray}
The variables with a circle or a tilde are even and $O(1)$ at the
center, while the variables with a hat are odd and $O(x)$. The
perturbed spacetime remains CSS if and only if all seven perturbation
variables are independent of $\tau$. These seven variables obey
evolution equations without any constraints, except for the trivial
ones that arise when one writes a wave equation in first-order form,
given below in (\ref{chiconstraint},\ref{kconstraint}). Applying the
same rescaling to the source terms $S_1$ to $S_5$, we obtain, in the
notation $Bu\equiv S(u)$, the source terms in the evolution equations
for $\tilde\chi$ to $\overcirc\psi$. To these we add two evolution
equations $\overcirc\chi$ and $\overcirc k$, which have only source
terms, but no $x$-derivatives. We obtain them by solving
(\ref{defdot},\ref{defprime}) for $f_{,\tau}$ in terms of $\dot f$ and
$f'$. Putting all seven source terms together, we have
\begin{eqnarray}
S(\tilde\chi) &=& {a\sqrt{1-V^2}\over g} 
{S_1+VS_2+2V(\bar\mu-\bar U)(sx)^{-1} S_5 \over 1-V^2} -(l+3)\tilde\chi, \\
S(\hat\chi) &=& {a\sqrt{1-V^2}\over g} 
{VS_1+S_2+2V^2(\bar\mu-\bar U)(sx)^{-1} S_5 \over 1-V^2} -(l+3)\hat\chi, \\
S(\tilde k) &=& {a\sqrt{1-V^2}\over g} 
{S_3+\kappa VS_4-2V\kappa \bar Usx S_5 \over 1-\kappa V^2} -(l+1)\tilde k \\
S(\hat k) &=& {a\sqrt{1-V^2}\over g} 
{VS_3+S_4-2V^2 \kappa \bar Usx S_5 \over 1-\kappa V^2} -(l+1)\hat k, \\
S(\overcirc\psi) &=& {a\sqrt{1-V^2}\over g} 
S_5 - (l+1)\overcirc\psi, \\
S(\overcirc\chi) &=& {as\over\sqrt{1-V^2}}\left[\left(Vx+{1\over
sg}\right) \tilde\chi - \left(x+{V\over sg}\right)\hat\chi\right] -
(l+2)\overcirc\chi, \\
S(\overcirc k) &=& {as\over\sqrt{1-V^2}}\left[\left(Vx+{1\over
sg}\right) \tilde k - \left(x+{V\over sg}\right)\hat k\right] -
l\overcirc k
\end{eqnarray}
Finally, solving (\ref{defdot},\ref{defprime}) for $f_{,x}$ in terms of
$\dot f$ and $f'$, we obtain two constraint equations, that is,
equations that do not contain $\tau$-derivatives. They are
\begin{eqnarray}
\label{chiconstraint}
\overcirc\chi_{,x} &=& {as\over\sqrt{1-V^2}}(\hat\chi-V\tilde\chi), \\
\label{kconstraint}
\overcirc k_{,x} &=& {as\over\sqrt{1-V^2}}(\hat k-V\tilde k).
\end{eqnarray}
The intermediate source terms $S_i$ are
\begin{eqnarray}
S_1&=& -\left[3\bar\mu+2(l+2)\bar U\right]\tilde\chi
+\left[5\bar\nu+2(l+1)\bar W\right]\hat\chi 
+4(\bar U-\bar\mu)(sx)^{-2}\tilde k \nonumber \\
&&+2\left[(l-1)(\bar\mu -\bar U)\bar W+2\bar\mu\bar\nu
+e^{-2\tau}(\mu'-\dot\nu)\right](sx)^{-1}\overcirc\psi
\nonumber \\
&&+\left[-2l\bar U^2 + (4l+8)\bar\nu\bar W - (2l+8)\bar\mu\bar U
- (2l^2+4)(sx)^{-2} {m\over r} + [-(1-\kappa)l+2\kappa+2] \bar\rho
+ 4\bar\nu^2 \right] \overcirc\chi \nonumber \\
&& +4\left[\bar\nu^2 +\left(\bar\rho-3(sx)^{-2}{m\over r}\right)
+(l+1)(\bar U-\bar\mu)\bar U\right](sx)^{-2}\overcirc k, \\
S_2&=&\bar\nu\tilde\chi-\bar\mu\hat\chi, \\
S_3&=&-(1+\kappa )\bar U(sx)^2\tilde\chi 
+ (1-\kappa )\bar W(sx)^2\hat\chi
-\left[(4+2\kappa+2l)\bar U+\kappa \bar\mu\right]\tilde k
+\left[\bar\nu+2\kappa(l+1)\bar W\right]\hat k \nonumber\\
&&+2\left[(1-\kappa )\bar\mu\bar W-(1+\kappa )\bar\nu\bar U-(\kappa
l+ 2\kappa+1)\bar U\bar W\right](sx)\overcirc\psi \nonumber \\
&&+\left\{-\left[(1+\kappa)l+2+4\kappa\right](sx)^2\bar U^2
+\left[(1-\kappa)l-2\kappa\right](sx)^2\bar W^2
+{1-\kappa\over 2}(l^2+l+2)
-4\kappa(sx)^2\bar\mu\bar U +4\kappa(sx)^2\bar\rho
\right\}\overcirc\chi \nonumber \\
&&+\left\{\left[-(1-\kappa)l^2-(\kappa+3)l-2-2\kappa\right]\bar U^2
-\left[2\kappa l^2-(1-\kappa)l-4\right](sx)^{-2}{m\over r}
-4\kappa\bar\mu\bar U + 4 \kappa\bar\rho
\right\} \overcirc k, \\
S_4&=&\bar\nu\tilde k-\bar\mu\hat k, \\ 
S_5&=&-(sx)\hat\chi -\left[(l+1)\bar U+2\bar\mu\right]\overcirc\psi 
-[(l+2)\bar W+2\bar\nu](sx)\overcirc\chi
-2\bar\nu(sx)^{-1}\overcirc k.
\end{eqnarray}
We have used the background equations of motion, and have introduced
the new coefficient $e^{-2\tau}(\mu'-\dot\nu)$, which is again
independent of $\tau$ on a CSS background.  Note that all coefficients
in the evolution equations are explicitly regular when one takes into
account that $\bar U-\bar \mu=O(x^2)$, $\bar\rho-(m/r)/(sx)^2=O(x^2)$
and $e^{-2\tau}(\mu'-\dot\nu)=O(x)$. The two terms $2(l+1)\bar
W\hat\chi$ and $2\kappa(l+1)\bar W\hat k$ are the equivalent of the term
$2(l+1)\phi'/r$ of the toy model wave equation that we discussed
above. They can be regularised in the same way.

\section{A toy model for the evolution equations}
\label{section:toymodel}

We have tested various numerical methods on a toy model, the scalar
wave equation in flat spacetime. This model is also useful as an
illustration of the types of variable and the methods we use. 

Let $\Phi$ obey the free wave equation on the flat spacetime
$ds^2=-dt^2+dr^2+r^2d\Omega^2$. We make the ansatz
\begin{equation}
\Phi=\sum_{l,m} \phi_{lm}(r,t)
Y_{lm}(\theta,\varphi). 
\end{equation}
We now consider a particular value of $l$ and $m$, and drop these
suffixes. Then $\phi_{lm}$ obeys
\begin{equation}
\label{flatspace_wave}
-\phi_{,tt} + \phi_{,rr} + {2\over r} \phi_{,r} - {l(l+1)\over r^2}\phi = 0. 
\end{equation}
As in \cite{fluidpert} we introduce the rescaled and first-order
variables
\begin{equation}
\bar\phi \equiv r^{-l} \phi, \qquad  \dot{\bar\phi} \equiv
\bar\phi_{,t}, \qquad {\bar\phi}' \equiv \bar\phi_{,r}.
\end{equation}
$\Phi(t,r,\theta,\varphi)=\Phi(t,x,y,z)$ is analytic in Cartesian
coordinates at the origin if and only if $\phi$ is analytic in $r$
with only even powers of $r$. In this sense $\bar\phi$ and
$\dot{\bar\phi}$ are even functions of $r$, and are generically finite
and nonzero at $r=0$, and ${\bar\phi}'$ is odd and generically $O(r)$
at $r=0$.

We now go over to self-similarity coordinates $x$ and $\tau$ defined
in (\ref{xtau}) (with $s=1$). In order to mimic the fluid
perturbation equations, we rescale the first-order
variables once again as
\begin{equation}
\overcirc\phi \equiv  e^{-n\tau}\bar\phi, \qquad
\tilde\phi \equiv  e^{-(n+1)\tau} \dot{\bar\phi}, \qquad
\hat\phi \equiv e^{-(n+1)\tau} \bar\phi'.
\end{equation}
The most natural choice of $n$ for the toy model is $n=0$, but for the
fluid perturbations, it will be fixed by the requirement that the
perturbed spacetime remains self-similar if the perturbation variables
do not grow or decay with $\tau$. We therefore leave it free. We
finally obtain the first-order system
\begin{eqnarray}
\tilde\phi_{,\tau} &=& -x \tilde\phi_{,x} + \hat\phi_{,x} + 2(l+1)
{\hat\phi\over x} - (n+1) \tilde\phi, \\
\hat\phi_{,\tau} &=& \tilde\phi_{,x} - x \hat\phi_{,x} -  (n+1) \hat\phi, \\
\overcirc\phi_{,\tau} &=& \tilde\phi -x\hat\phi - n\overcirc\phi.
\end{eqnarray}
There is also the trivial constraint
\begin{equation}
\label{trivial_constraint}
\overcirc\phi_{,x} = \hat\phi
\end{equation}
that follows from the definition of the first-order variable
$\hat\phi$.  

One significant property that the toy model has in common with the
full perturbation equations is that $\overcirc\phi$ and $\tilde\phi$
are even and generically nonzero at $x=0$ while $\hat\phi$ is odd and
generically $O(x)$. All terms except $\hat\phi/x$ are explicitly
regular, and do not require special treatment at the origin. In
particular, there is no term of the type $\phi/x^2$ left. Experience
shows that terms of the form $\hat\phi/x$, where $\hat\phi$ is an odd
function of $x$ and $O(x)$, can be regularised without giving rise to
instabilities, while division by $x^2$ is much more troublesome.

In other, unimportant, aspects the toy model is simpler than the full
perturbation equations. As there is no fluid frame in this toy model,
we choose the dot and prime to be frame derivatives with respect to
the constant $r$ frame. As the spacetime is flat, the dot and prime
derivatives commute. Finally, in the toy model $\overcirc\phi$ does
not couple back to $\hat\phi$ and $\tilde\phi$, and therefore plays
only a passive role.

\section{Finite differencing}
\label{appendix:methods}

For the evolution equations (\ref{evolution_equations}) at hand, the
eigenvalues and eigenvectors of the matrix $A$ can be calculated in
closed form. The eigenvalues of $A$ are $dx/d\tau$ on characteristics
of the equations: fluid world lines, radial light rays, and radial
matter (sound wave) characteristics. The characteristics are symmetric
around the line $x=0$, but with increasing $x$ they tip over until at
sufficiently large $x$ all eigenvalues are negative. This means that
at large $x$ information travels only from smaller to larger $x$. The
reason for this is of course that while $x=0$ and lines of small
constant $x$ are timelike, lines of large constant $x$ are
spacelike. The ``outer boundary'' $x=x_{\rm max}$ of our numerical
domain $0\le x\le x_{\rm max}$ is therefore a future spacelike
boundary, and so no boundary condition is required there.

In order to use it to obtain a free boundary condition, we make the
numerical method reflect the propagation of information, so that at
$x=x_{\rm max}$ all $x$-derivatives are calculated using one-sided
finite differences. Following \cite{Leveque}, we split $A$ into a left
and a right-moving part. Let $V$ be the matrix of (column)
eigenvectors of $A$. Let $\Lambda$ be the diagonal matrix composed of
the corresponding eigenvalues. Then $A=V \Lambda V^{-1}$. Let
$\Lambda_+$ be $\Lambda$ with zeros in the place of the negative
eigenvalues, and let $\Lambda_-$ be $\Lambda$ with zeros in the place
of the positive eigenvalues. Then define $A_\pm =V \Lambda_\pm
V^{-1}$. It is clear that $A=A_++A_-$. We now use $A_+$ with left
derivatives and $A_-$ with right derivatives.
\begin{equation}
\label{predictor}
{du^n_j\over d\tau} =  ( A_+ D_+u^n_j + A_- D_-u^n_j + B u^n_j ),
\end{equation}
with the one-sided derivatives
\begin{equation}
D_+u^n_j = {u^n_{j+1} - u^n_j \over \Delta x}, \qquad
D_-u^n_j = {u^n_j - u^n_{j-1} \over \Delta x}.
\end{equation}
Here the coefficient matrices $A_\pm$, $B$ and $C$ are evaluated at
$x_j$. After each time step, the constrained variables $w$ are
obtained at $\tau^{n+1}$ from the $u^{n+1}$.  This scheme is
first-order accurate in $x$. (Note that we have not discretized in
$\tau$ yet.) Left differences at $x=0$ are evaluated using ghost
points based on the fact that all grid functions $u$ are either even
or odd in $x$. This scheme can be thought of as the Godunov method
applied to a linear equation, and we shall refer to it as the
first-order Godunov scheme, or GD1.
To obtain a second independent finite differencing scheme, we have also
used the second-order one-sided derivatives
\begin{equation}
D_+u^n_j = {4u^n_{j+1} - 3u^n_j - u^n_{j+2} \over 2\Delta x}, \qquad
D_-u^n_j = -{4u^n_{j-1} - 3u^n_j - u^n_{j-2} \over 2\Delta x}.
\end{equation}
We shall refer to this scheme as GD2. At the outer boundary $x=x_{\rm
max}$, no right derivatives are required, as there all information
travels from the left to the right. ($A_+$ vanishes.) All variables
$u$ are either even or odd functions of $x$. At the inner boundary
$x=0$, left derivatives are calculated using fictitious grid points at
negative $x$ that are obtained as $u(-x)=\pm u(x)$.
By construction, the Godunov method has the advantage that it does not
require a special outer boundary conditions. Unexpectedly, it also has
the advantage that it handles terms of the form $\hat\phi/x$ term at
the center without any special treatment. It is also completely free from
high-frequency grid-modes. This last property is less surprising when
one thinks of GD1 as centered differencing plus a dissipative term
\cite{Leveque}. 

We have also implemented a more standard finite differencing scheme
based on centered differences in $x$, with a (first or second order)
one-sided derivative at the outer boundary $x=x_{\rm max}$.  The
center is again handled by ghost points, using $u(-x)=\pm u(x)$.
Using naive centered differences, the code is unstable at the center
because of the source term $2(l+1)\hat\phi/x$ in the toy model, and
similar terms in the perturbation equations. A well-known remedy is to
include this source term into the transport terms in the way suggested
by the identity
\begin{equation}
\label{centertrick}
\hat\phi_{,x} + {2(l+1)\over x}\hat\phi = (2l+3) {\partial
(x^{2l+2}\hat\phi) \over \partial (x^{2l+3})}.
\end{equation}
In the toy model wave equation this procedure slows down the central
instability enough so that it can be suppressed by numerical
viscosity. We have added a centered difference expression for
$u_{,\tau}=c u_{,xx}+\dots$, with $c$ of the order of
$10^{-3}$. However, numerical viscosity falsifies the results too much
both in the toy model and in the actual problem, and has therefore not
been used in any of the results of this paper. We shall refer to
centered differencing with (\ref{centertrick}) as CD3, and without as
CD4. 

\section{Weak solutions}
\label{appendix:weakmodes}

We now discuss a problem that affects the fluid and metric
perturbations that we want to investigate, but that is already present
and more easily understood in the toy model.

The spacetime point $r=t=0$ is singled out in the self-similarity coordinates
$x$ and $\tau$, but it is not preferred on the flat background. A
solution arising from generic $C^2$ initial data at
$t=t_0<0$ has finite $t$ and $r$ derivatives at $t=r=0$. The
coordinates $x$ and $\tau$, however, have been designed to ``zoom in''
on this spacetime point: we are looking at a smooth solution on ever
smaller scales. Therefore, as $\tau\to\infty$, a generic solution of
the toy model wave equation should behave at large $\tau$ as
\begin{equation}
\label{generic_falloff}
\overcirc\phi\to \bar\phi(0,0)e^{-n\tau}, \quad
\tilde\phi\to \bar\phi_{,t}(0,0)e^{-(n+1)\tau}, \quad
\hat\phi\to {1\over 2}\bar\phi_{,rr}(0,0)e^{-(n+2)\tau}.
\end{equation}
In deriving this fall-off we have assumed that $\phi$ is at least
twice differentiable everywhere. But as $x=1$ is a characteristic of
the wave equation, $\phi$ and its derivatives are allowed to be
discontinuous there. In particular, data on $x> 1$ do not influence
the solution on $x\le 1$. Therefore, solutions exist that vanish on
$x\le 1$ for all $\tau$ but not for $x>1$. In particular, making a
power-series ansatz for the region $x>1$,
\begin{equation}
\tilde\phi(x,\tau) = e^{\lambda\tau} \left[\tilde\phi_1 (x-1) +
\tilde\phi_2 (x-1)^2 + \dots\right], \qquad
\hat\phi(x,\tau) = e^{\lambda\tau} \left[\hat\phi_1 (x-1) +
\hat\phi_2 (x-1)^2 + \dots\right],
\end{equation}
we find a formal solution with $\lambda=l-1-n$, which is slower than
the expected falloff by $l$. We believe that a corresponding weak
solution of the wave equation exists that is analytic for $x>1$ (and
vanishes for $x\le 1$), but have not proved it. A finite difference
counterpart certainly does exist, and for GD1 and GD2 it falls off (or
grows, depending on $n$ and $l$) with the calculated value of
$\lambda$. It therefore dominates over the generic smooth solution at
large $\tau$. We want to exclude it in studies of critical collapse,
as we are interested only in the time evolution of smooth perturbation
initial data. Numerical schemes, however, do not distinguish between
smooth and unsmooth data, and this solution turns up in some of them,
hiding the everywhere $C^2$ solutions we are interested in. In the
future we shall refer to these as ``weak solutions at the light cone''
or simply ``weak modes''. We should stress that, depending on the
finite differencing scheme, weak solutions of the finite difference
equations may or may not converge to weak solutions of the continuum
equations, but they are always there.

We find that both Godunov schemes when applied to the toy model wave
equations develop weak solutions that are continuous but not
differentiable at $x=1$. For GD1, the finite difference mode
corresponding to the weak solution can be analyzed easily. $u_{,\tau}$
at the first grid point with $x>1$ depends only on $u$ at that point,
and on the next point to the left -- but there all fields vanish in a
weak solution. A calculation shows that $u$ at the first grid point
with $x>1$ depends exponentially on $\tau$. The exponent is either
$\lambda=l-1-n+O(\Delta x)$, which is the continuum weak solution, or
$\lambda = -2/\Delta x + O(1)$, which is a finite differencing
artifact that decays quickly. 

In GD1, the dominant weak is clearly not differentiable at $x=1$, and
this also shows up in independent residual evaluation as a sharp
peak. If the last grid point is exactly $x=1$, the numerical domain
can consistently be truncated there, and no weak modes can arise.
Because of its wider stencil GD2 is not completely causal, and so part
of the weak mode propagates to $x<1$. The numerical equivalent of the
weak mode therefore appears differentiable at $x=1$, but there is
still a peak there in independent residual evaluation, although less
sharp.  If we attempt to truncate GD2 at $x=1$, we need to introduce
an explicit boundary condition there. Doing this by using a
first-order right derivative, or a centered derivative at $x=1-\Delta
x$ introduces spurious reflections at the boundary. Without
truncation, we have tried updating the grid point $x=1+\Delta x$ by
interpolation from the neighboring grid points. This does not suppress
the weak mode. Updating the three points $x=1$, $x=1+\Delta x$ and
$x=1+2\Delta x$ by interpolation results in a different spurious
mode. This holds both for GD1 and GD2.

\section{Imposing the constraints}
\label{appendix:constraints}

Some perturbation equations contain only $x$-derivatives, and so are
constraints rather than evolution equations. Even the free wave
equation has such a constraint when it is written in first-order
form. Constraints of the form (\ref{trivial_constraint}) are solved by
integration from the center out, using the trapezoid rule,
\begin{equation}
\label{trapezoid}
\overcirc\phi^{i+1} = \overcirc\phi^i + {\Delta x\over
2}\left(\hat\phi^{i+1} + \hat\phi^i\right),
\end{equation} 
where the starting point $\overcirc\phi$ at the center is determined by the
evolution equation. This method is second-order accurate. We also use
the exact of these finite difference equations to obtain $\hat\phi$
from $\overcirc\phi$ by differentiation.

Time evolution commutes with the constraints in the continuum limit,
but in the discretized equations this is only approximately true. In
order to find the modes, we need a matrix $T$ that acts only on a set
of functions $u$ that can be freely specified in the initial data and
is therefore of full rank. We start with a larger matrix $T'$ and have
to reduce it using a numerical solution of the constraints. This
reduction, however, is not unique, and although different reductions
are equivalent in the continuum equations, they are not in the finite
difference equations.

We discuss these issues in the toy model. The matrix $T'$ acts on
$\overcirc\phi$, $\hat\phi$ and $\tilde\phi$, so that $T'$ is a
$(3N)^2$ matrix. There are two natural choices for the free variables
on which $T$ acts: either $\overcirc\phi$ and $\tilde\phi$, or the
functions $\hat\phi$ and $\tilde\phi$ plus the number
$\overcirc\phi(0)$. Note that $\hat\phi(0)=0$ by definition, so that
in either case $T$ is a $(2N)^2$ matrix.

We denote the constraint solution scheme (\ref{trapezoid}) by $I$ (for
integration) and its numerical inverse by $D$ (for differentiation). 
In loose matrix notation we can then write the first possibility as
\begin{equation}
u=\left(\begin{array}{c} \tilde\phi \\\overcirc\phi
 \end{array}\right),
\qquad T_1\equiv\left(\begin{array}{ccc}
1&0&0\\
0&0&1
\end{array}\right) T'
\left(\begin{array}{cc}
1&0\\
0&D\\
0&1
\end{array}\right),
\end{equation}
and 
\begin{equation}
u=\left(\begin{array}{c} \tilde\phi \\\hat\phi
 \end{array}\right),
\qquad T_2\equiv\left(\begin{array}{ccc}
1&0&0\\
0&1&0
\end{array}\right) T'
\left(\begin{array}{cc}
1&0\\
0&1\\
0&I
\end{array}\right).
\end{equation}
We find that $T_2$ works much better than $T_1$, in having grid
modes. This is not surprising giving that integration has a smoothing
property. An alternative to $T_1$ is to re-impose the constraints by
integration after evolution, 
\begin{equation}
u=\left(\begin{array}{c} \tilde\phi \\\overcirc\phi
 \end{array}\right),
\qquad T_3\equiv\left(\begin{array}{ccc}
1&0&0\\
0&I&0
\end{array}\right) T'
\left(\begin{array}{cc}
1&0\\
0&D\\
0&0
\end{array}\right).
\end{equation}
But the matrix $T_3$ is similar to the matrix $T_2$, 
\begin{equation}
T_2=\left(\begin{array}{cc}
1&0\\
0&D
\end{array}\right) T_3
\left(\begin{array}{cc}
1&0\\
0&I
\end{array}\right),
\end{equation}
and so $T_2$ and $T_3$ have the same eigenvalues, although of course
not the same eigenvectors. This is borne out numerically up to small
rounding errors. 

In the spherical and the $l=1$ polar perturbations, nontrivial
constraint equations arise which are of the form
\begin{equation}
xu_{,x} = M(x)u+s,
\end{equation}
where $u$ stands for a subset of variables, and the source terms
$s$ are linear in the other variables, already known at this time
level. The coefficient matrix $M(x)$, an even function of $x$, is
typically nonzero at the origin. The ODE is therefore singular at
$x=0$. We look for solutions $u(x)$ that are even regular functions of
$x$. These obey $M(0)u(0)=s(0)$. From this boundary condition, the
ODEs are solved by the implicit, second-order accurate scheme
\begin{equation}
y_{i+1} = \left(1 - \epsilon M_{i+1}\right)^{-1} 
\left[ \left(1 + \epsilon M_i\right)y_i 
+ \epsilon \left(s_i + s_{i+1}\right)
\right], \qquad \epsilon\equiv (x_{i+1}-x_i)/(x_{i+1}+x_i).
\end{equation}




\begin{references}

\bibitem{Choptuik} M. W. Choptuik, Phys. Rev. Lett. {\bf 70}, 9 (1993). 

\bibitem{critreview} C. Gundlach, Living Reviews {\bf 1999}-4,
published electronically as http://www.livingreviews.org. See also
Physics Reports, in preparation.

\bibitem{KHA1} T. Koike, T. Hara, and S. Adachi, {
Phys. Rev. Lett.} {\bf 74}, 5170 (1995). 

\bibitem{EvansColeman} C. R. Evans and J. S. Coleman, {
Phys. Rev. Lett.} {\bf 72}, 1782 (1994).

\bibitem{Maison} D. Maison, { Phys. Lett.} B {\bf 366}, 82 (1996).

\bibitem{KHA2} T. Koike, T. Hara, and S. Adachi, 
Phys. Rev. D {\bf 59}, 104008 (1999). 

\bibitem{NeilsenChoptuik} D. W. Neilsen and M. W. Choptuik,
Class. Quant. Grav. {\bf 17}, 761 (2000).

\bibitem{codim} C. Gundlach,  Phys. Rev. D {\bf 57}, R7075 (1998).

\bibitem{fluidpert} C. Gundlach and J. M. Mart\'\i n-Garc\'\i a,
Phys. Rev. D. {\bf 61}, 084024 (2000).

\bibitem{critscalar} J. M. Mart\'\i n-Garc\'\i a and C. Gundlach,
Phys. Rev. D {\bf 59}, 064031 (1999).

\bibitem{angmom} C. Gundlach, Phys. Rev. D {\bf 57}, R7080 (1998).

\bibitem{Gundlach_Chop2} C. Gundlach, { Phys. Rev.} D {\bf 55}, 695
(1997).

\bibitem{scalingfunctions} C. Gundlach, ``Critical gravitational
collapse with angular momentum: from critical exponents to scaling
functions'', in preparation.

\bibitem{Leveque} R. J. LeVeque, {\it Numerical methods for
conservation laws}, Birkh\"auser, Basel 1992.

\end{references}
\end{document}